\documentclass[12pt]{JHEP3}
\pdfoutput=1
\usepackage{amsfonts}
\usepackage{amscd}
\usepackage{amssymb}
\usepackage{amsmath}
\usepackage{epsfig}
\usepackage{latexsym}

\def \be  {\begin{equation}}
\def \ee  {\end{equation}}
\def \ba  {\begin{eqnarray}}
\def \ea  {\end{eqnarray}}
\def \bb  {\begin{thebibliography}}
\def \eb  {\end{thebibliography}}
\def \lab #1 {\label{#1}}

\newcommand\cA{\mathcal{A}}

\newcommand\cD{\mathbb{D}}

\newcommand\cO{\mathcal{O}}

\newcommand\cN{\mathcal{N}}
\newcommand\C {\mathbb{C }}

\newcommand\CP {\mathbb{CP}}

\newcommand\rd{\mathrm{d}}

\newcommand\im{\mathrm{i}}
\newcommand\la{\langle}
\newcommand\ra{\rangle}
\newcommand\del{\partial}
\newcommand\delbar{\bar{\partial}}
\newcommand\Dbar{\bar{\mathcal{D}}}
\newcommand\tr{\mathrm{Tr}}
\newcommand\MHVbar{\overline{\mbox{MHV}}}

\newcommand\CSW{*}

\title{The Complete Planar S-matrix of $\cN=4$ SYM 
	as a Wilson Loop in Twistor Space}

\author{Lionel Mason\\
	The Mathematical Institute,\\
	24-29 St.~Giles', Oxford, OX1 3LB,
	United~Kingdom}

\author{David Skinner\\
	Perimeter Institute for Theoretical Physics,\\ 
	31~Caroline~St., Waterloo, ON, N2L 2Y5, 
	Canada}

\abstract{
We show that the complete planar S-matrix of $\cN=4$ super Yang-Mills -- including all N$^k$MHV partial amplitudes to all loops -- is equivalent to the correlation function of a supersymmetric Wilson loop in twistor space.  Remarkably, the entire classical S-matrix arises from evaluating the correlation function in the self-dual sector, while the expansion of the correlation function in powers of the Yang-Mills coupling constant provides the loop expansion of the amplitudes. We support our proposal with explicit computations of the $n$ particle NMHV and N$^2$MHV trees, the integrands of the 1-loop MHV and NMHV amplitudes, and the $n$ particle 2-loop MHV amplitude. These calculations are performed using the twistor action in axial gauge. In this gauge, the Feynman diagrams of the correlation function are the planar duals of the usual MHV diagrams for the scattering amplitude. The results are presented in the form of a sum of products of dual superconformal invariants in (momentum) twistor space, and agree with the expressions derived in the companion paper~\cite{CSWMat} directly from the MHV rules. The twistor space Wilson loop is a natural supersymmetric generalization of the standard Wilson loop used to compute MHV amplitudes. We show how the Penrose-Ward transform can be used to determine a corresponding supersymmetrization on space-time and give the corresponding superconnection in the abelian case.

}  

\begin{document}


\section{Introduction}
\label{sec:intro}

The main purpose of this paper is to show that all $n$-particle planar amplitudes in $\cN=4$ super Yang-Mills are captured by the correlation function of a supersymmetric Wilson loop 
\be
	\big\la {\rm W}[C_n]\big\ra
\label{main}
\ee
in (momentum) twistor space.  This includes both trees and all loop corrections for all N$^k$MHV partial amplitudes, up to an overall factor of the MHV tree.

We defer the full explanation of this operator to section~\ref{sec:WL}, but in brief it is a supersymmetric generalization of the translation into twistor space of a space-time Wilson loop associated to a null polygon.  The Wilson loop W$[C_n]$ is a natural holomorphic analogue of the standard space-time definition of the trace of the holonomy of a connection around a loop: it computes the trace of the holonomy of the process of finding holomorphic frames of an almost complex  bundle around a nodal curve $C_n$ in supertwistor space $\CP^{3|4}$. This curve is a polygon whose `edges' are Riemann spheres (complex projective lines) dual to the vertices of an $n$-sided null polygon $C'_n$ in dual conformal space-time (see figure~\ref{fig:polygons}).   For much of the discussion here, all that is needed of the geometry of the twistor correspondence is that points in space-time correspond to complex projective lines ($\CP^1$s) in twistor space, in such a way that intersection of lines in twistor space is equivalent to the points lying on a null geodesic. See {\it e.g.}~\cite{HuggettTod,Penrose:1986ca,WardWells} for introductions to twistor geometry.

\FIGURE[t]{
	\includegraphics[height=40mm]{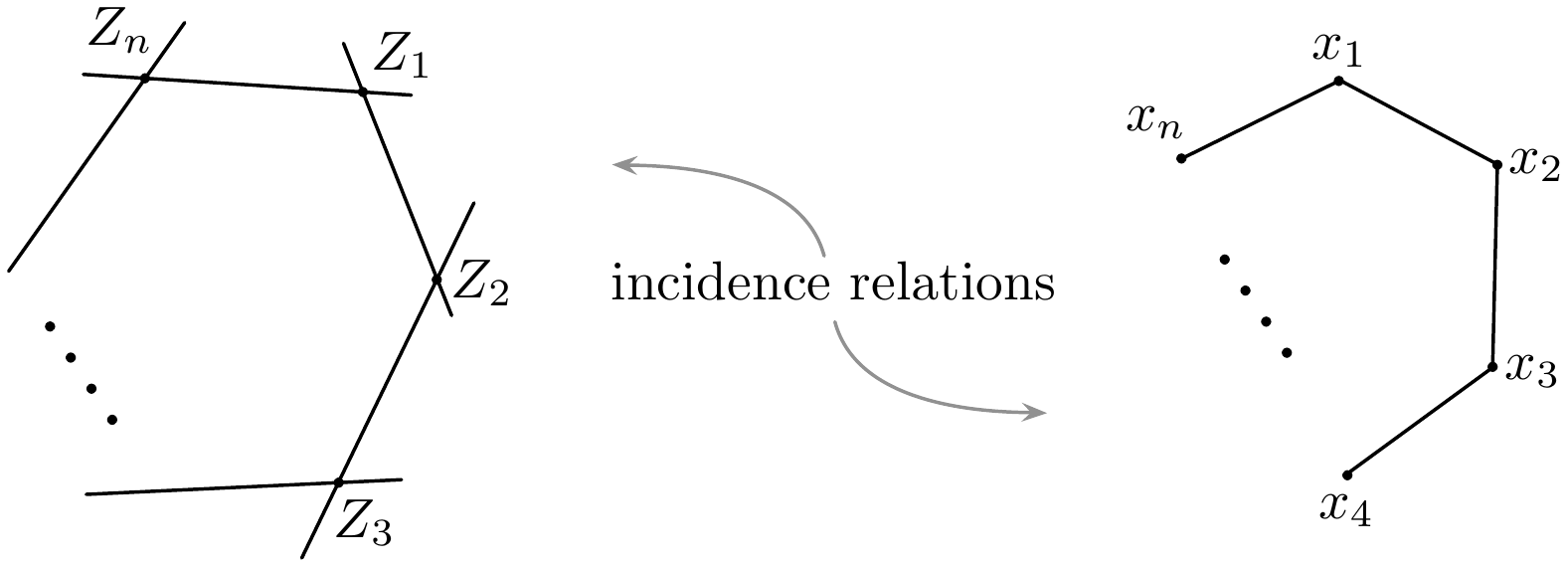}
	\caption{The nodal curve $C_n$ in twistor space that corresponds to the null polygon in space-time. The
	twistors $Z_i$ represent the edges of the space-time polygon, while the lines $(Z_i,Z_{i-1})$ correspond to 
	the space-time vertices $x_i$.}
	\label{fig:polygons}
}

The relation to scattering amplitudes comes from treating the twistor space to be the space of the fundamental representation of the \emph{dual} superconformal group~\cite{Drummond:2008bq}, rather than the usual superconformal group. This dual group acts on the space of `region momenta' of a scattering amplitude -- defined up to translation by $x_i-x_{i+1} = p_i$ where $p_i$ is the momentum of the $i^{\rm th}$ particle in the scattering process -- in exactly the same way as the usual superconformal group acts on space-time. The twistor space associated to region momentum space is called \emph{momentum} twistor space; it has the same relation to dual space-time as ordinary twistor space has to usual space-time. (See {\it e.g.}~\cite{Hodges:2009hk,Mason:2009qx} for more detailed introductions to momentum twistors.)

Our proposal is motivated by the corresponding proposal in dual space-time that null polygonal Wilson loops yield the planar MHV amplitude to all loop orders~\cite{Drummond:2007aua,Drummond:2007au,Drummond:2008aq,Brandhuber:2007yx, Anastasiou:2009kna,DelDuca:2009au,DelDuca:2010zg,DelDuca:2010zp,Goncharov:2010jf}.  This was in turn motivated by the proposal that taking the polygon to bound a minimal surface in AdS$_5$, the area of this surface gives the strong coupling limit of the same planar MHV amplitude~\cite{Alday:2007hr,Alday:2009ga,Alday:2009dv,Alday:2010vh}. However, on dual space-time it has not yet been possible obtain non MHV amplitudes in this way.

To support our main conjecture, we use the correlation function~\eqref{main} to obtain the NMHV and N$^2$MHV tree amplitudes, and the integrands of the MHV and NMHV 1-loop and MHV 2-loop amplitudes, each for an arbitrary number of particles.   These calculations turn out to be straightforward using the twistor action for $\cN=4$ SYM that was introduced in~\cite{Mason:2005zm,Boels:2006ir,Boels:2007qn}.  The twistor action is reviewed in section~\ref{sec:action}. It consists of a holomorphic Chern-Simons theory, corresponding to self-dual $\cN=4$ SYM,  together with an infinite series of MHV vertices that are supported on lines in twistor space. Since it lives on a six (real) dimensional space, the twistor action has a larger amount of gauge freedom than the usual space-time action, providing access to gauges not available on space-time. In particular, in an axial gauge, its Feynman diagrams are the MHV diagrams of~\cite{Cachazo:2004kj}, here being used to compute a correlation function rather than a scattering amplitude.

In the companion paper~\cite{CSWMat}, we translated the MHV formalism for scattering amplitudes into momentum twistor space. There, we found that the vertices of an MHV diagram are represented simply by 1 on momentum twistor space, while each propagator corresponds to an R invariant~\cite{Drummond:2008bq}, constructed from the pair of twistors associated to each of the two regions separated by the propagator, and an additional reference twistor. Remarkably,  the diagrams generated by the correlator~\eqref{main} are precisely the planar dual graphs of the MHV diagrams for the corresponding planar amplitude!  In particular, the number $V$ of MHV vertices in the Feynman diagrams of the correlation function~\eqref{main} is simply the loop order $\ell$ of the scattering process.  Consequently, the entire classical S-matrix of $\cN=4$ SYM comes from the correlation function of the supersymmetric Wilson loop operator purely in the holomorphic Chern-Simons theory, as we will see in section~\ref{sec:trees}.

In section~\ref{sec:loops} we consider corrections to the correlator~\eqref{main} in powers of the Yang-Mills coupling. The fact that the diagrams of the twistor Wilson loop~\eqref{main} are dual to the usual MHV diagram, with the expression agreeing identically even before performing the loop integrals, makes it obvious that the Wilson loop corresponds to the scattering amplitude, at least at a computational level. This is in marked contrast to the space-time calculation, where the relation between the Wilson loop and scattering amplitudes is mysterious, with seemingly very different constructions providing identical answers at the final stage (compare~\cite{Drummond:2008aq} and~\cite{Bern:2008ap}, for example).

Using the twistor action naturally packages the result of the correlator as the same sum of products of R invariants as found in~\cite{CSWMat} from the MHV formalism for scattering amplitudes. To compute the loop amplitudes, one must integrate these R invariants over copies of the twistor space, two for each loop. Loop integrals over twistor space are often tractable, and have been explicitly carried out in various 1- and 2-loop cases~\cite{Hodges:2010kq,Mason:2010pg,Drummond:2010mb,Alday:2010jz} using the Coulomb branch regulator introduced in~\cite{Alday:2009zm} to treat the infra-red divergences. However, the \emph{integrand} of the loop amplitude -- a concept that makes sense only in the planar limit where one may use the cyclic ordering to define a loop momentum common to different diagrams -- is an interesting object in its own right, and has recently been proved~\cite{Arkani-Hamed:2010kv} to possess the full Yangian invariance of planar $\cN=4$ SYM, first discovered in scattering amplitudes in~\cite{Drummond:2009fd}. It is therefore intriguing that the correspondence between scattering amplitudes and the Wilson loop appears to hold even at the level of the integrand.

Finally, our twistor space Wilson loop is a supersymmetric formulation of a transcription of the standard space-time Wilson loop. Although the twistor formulation is especially elegant, the results of this paper strongly suggest that it is also possible to find a generalisation of the Wilson loop to all N$^k$MHV amplitudes purely on space-time. In section~\ref{sec:space-time} we explain how an off-shell version of the Penrose-Ward transform can be used to deduce the superconnection explicitly when it does correspond to one.  The trace of the holonomy of this superconnection should give rise to an equivalent space-time supersymmetric Wilson loop. In space-time, the twistor action most naturally corresponds to the Chalmers-Siegel action for $\cN=4$ SYM~\cite{Chalmers:1996rq,Siegel:1992za}. From this point of view, the MHV vertices that give rise to loop amplitudes essentially correspond to the Lagrangian insertions considered in~\cite{Alday:2010zy,Eden:2010zz,Eden:2010ce}.

\medskip

The ability to compute correlation functions using the twistor action is not at all limited to the specific operator in~\eqref{main}. The ideas of this paper will be presented in greater generality, with fuller explanations and further elaboration, in the forthcoming paper~\cite{withMat}.


\section{Twistor space and Wilson loops}
\label{sec:WL}

We begin by defining the operator in~\eqref{main}, explaining in general terms why this operator is expected to be related to a space-time Wilson loop.


\subsection{The off-shell Penrose-Ward correspondence}

To describe gauge theories on twistor space, we introduce a rank $N$ complex bundle $E$ over twistor space that has vanishing first Chern class and is equipped with a (0,1)-connection $\overline{D}=\bar\del+a$, where $a$ is a (0,1)-form with values in the Lie algebra of the complexification of the gauge group.  From $E$ we wish to construct a bundle $\widetilde{E}$ over some portion $U$ of conformally compactified, complexified space-time (say an affine patch of some real slice),  so our twistor bundle $(E,\overline{D})$ will be defined on the region $\widehat{U}\subset\CP^3$ swept out in twistor space by the lines X corresponding to points $x\in U$.

We do not assume that $\overline{D}^2=0$ on twistor space, and so our bundle has an {\em almost} complex structure.  Nevertheless, $\overline{D}^2$ necessarily does vanish when restricted to any ${\rm X}\cong\CP^1$, because $\overline{D}^2$ is a $(0,2)$-form.   Any complex bundle on $\CP^1$ that is topologically trivial is also holomorphically trivial for small enough $a$, so $E|_{\rm X}$ is holomorphically trivial.  Therefore we can find $N$ linearly independent global holomorphic sections of $E|_{\rm X}$ that by Liouville's theorem are unique up to a constant GL$(N)$ transformation.  The space $\Gamma(X, E|_{\rm X})$ of such sections is a copy of $\C^N$ associated to the point $x$ in space-time, and so forms the fibre of a space-time Yang-Mills bundle $\widetilde{E}$.

More explicitly, whenever $E|_{\rm X}$ is holomorphically trivial we can find a smooth gauge transformation $H(\lambda,\bar\lambda)$ such that $H^{-1}(\delbar+a)|_{\rm X} H = \delbar|_{\rm X}$, or equivalently
\be
	\left.a\right|_{\rm X} = -\delbar H\, H^{-1}\,.
\label{globalframe}
\ee
Here, $\lambda$ is a coordinate on the Riemann sphere X and the $\delbar$-operator in~\eqref{globalframe} acts in the $\bar\lambda$ direction.  As  the line X varies in $\widehat{U}$, the family $H(x,\lambda,\bar\lambda)$ of such gauge transformations  varies smoothly with $x\in U$.  The holomorphic frame $H:=H(x,\lambda,\bar\lambda)$ on X will play an important role in what follows, and exists even for an almost complex bundle $E$.  It is unique up to 
\be
	H(x,\lambda,\bar\lambda) \rightarrow H(x,\lambda,\bar\lambda)g(x) \, ,
\ee
where $g$ depends only on $x$ because it must be globally holomorphic in $\lambda$ to preserve~\eqref{globalframe}, and Liouville's theorem states that any globally holomorphic function on a Riemann sphere is constant.

Although we have constructed a space-time bundle from an arbitrary almost complex bundle $E$ on twistor space (subject only to c$_1(E)=0$ and the smallness of $a$), the $\overline{D}$-operator on twistor space (depending on six real dimensions) cannot in general be encoded in a connection $\nabla$ on four-dimensional space-time without imposing further conditions. Nevertheless, we will see that the concept of parallel propagation along null rays survives,  so even in the most general case we can define the holonomy around a null polygon, and hence the Wilson loop required for the amplitude discussion.  In certain circumstances, say for $U$ a Euclidean real slice~\cite{Mason:2005zm}, the condition for $\overline{D}$ to give rise to a generic off-shell connection on space-time can be characterised by the vanishing of some components of $\overline{D}^2$. (The vanishing of all components of $\overline{D}^2$ would imply that $\nabla$ is self-dual.) When the twistor $\overline{D}$-operator does determine a connection on space-time, the Wilson loop defined below agrees with that for the space-time connection.   We give an explicit demonstration of how this works in section~\ref{sec:space-time}.


\subsection{Parallel propagation and Wilson loops}

The standard Wilson loop for a curve $C$ in space-time is computed by first identifying the holonomy matrix obtained by parallel propagation of the fibre $\widetilde E_p$ of $\widetilde E$ at some basepoint $p\in C$ around $C$ and back to $p$ again; one then takes its trace.  The parallel propagation can be represented as the path-ordered exponential integral 
\be
	{\rm W}[A;C]=\tr\; {\rm P} \exp\left(-\int_C A\right) 
\ee
of the connection $A$ around $C$, and is a gauge invariant functional of both $A$ and $C$.

In the context of scattering amplitudes, we are particularly interested in parallel transport around a piecewise null polygon with vertices at points $x_i$ for $i=1,\ldots n$. The null ray segments correspond to the massless momenta in the scattering process via the formula
\be
	x_i-x_{i+1} = \lambda_i\bar\lambda_i,
\label{affineP}
\ee
so that the direction of the null ray is determined by the spinor $\lambda_i$. In twistor space, this null polygon corresponds to a nodal curve whose components are holomorphic lines X$_i$ (representing the vertices $x_i$) and whose nodes occur at locations $Z_i$ representing the null rays (see figure~\ref{fig:polygons}).

Parallel propagation along a null geodesic has a particularly straightforward interpretation in twistor space, because the null ray from $x_i$ to $x_{i+1}$ corresponds to a single point $Z_i$ in twistor space where the lines $X_i$ and $X_{i+1}$ intersect. The parallel propagator
\be
	{\rm U}(x_{i+1},x_i;\lambda_i) = {\rm P}\exp\left(-\int_{x_i}^{x_{i+1}} \!\!A\right)
\ee
from $x_i$ to $x_{i+1}$ tells us how to compare the fibre $\widetilde E_{x_i}$ with $\widetilde E_{x_{i+1}}$.  As we explained above, these space-time fibres are the spaces of global sections of the twistor bundle $E$ restricted to X$_i$ and X$_{i+1}$, respectively. But since ${\rm X}_i\cap{\rm X}_{i+1} = Z_i$, these sections can be compared directly at $Z_i$.  Thus parallel propagation along a null geodesic identifies the fibres of $\widetilde E$ along that geodesic with the fibre $E_{Z_i}$ of the twistor bundle.  In formul\ae, this is achieved by setting
\be
	{\rm U}(x_{i+1},x_i;\lambda_i) = H(x_{i+1},\lambda_i)^{-1} H(x_i,\lambda_i)\, ,
\label{parallelpropZ}
\ee
because $H(x,\lambda)$ is a holomorphic frame for $E$ restricted to X, so~\eqref{parallelpropZ} is a gauge transformation that compares the frames of the twistor bundle on the two Riemann spheres X$_i$ and X$_{i+1}$ at their intersection point $Z_i=(\lambda_i,\mu_i)$. The off-shell equivalence of these two forms of the parallel propagator will be proved in section~\ref{sec:space-time}.

Continuing all around the null polygon and taking the trace,  we find that the Wilson loop is represented on twistor space by 
\be
\begin{aligned}
	{\rm W}[C_n] &=\tr  \left({\rm P} \prod_{i=1}^n H(x_{i+1},\lambda_i)^{-1} H(x_{i},\lambda_i)\right)\\
	&=\tr  \left({\rm P} \prod_{i=1}^n H(x_i,\lambda_i)\, H(x_i,\lambda_{i-1})^{-1}\right)\,,
\end{aligned}
\label{WLtws1}
\ee
where ${\rm P}\prod$ denotes an ordered product, with $i$ increasing towards the left. The first line of~\eqref{WLtws1} follows directly from concatenating the parallel propagators~\eqref{parallelpropZ} to give the holonomy. The second line of~\eqref{WLtws1} re-partitions the product in terms of the holomorphic frames that live on each line. Recalling that $H(x_i,\lambda)$ was defined in~\eqref{globalframe} to satisfy
\be
	(\delbar+a)|_{{\rm X}_i}\,H=0\,,
\label{Hdef}
\ee
we see that $H(x_i,\lambda) H(x_i,\lambda_{i-1})^{-1}$ is the solution to~\eqref{Hdef} that obeys the boundary condition that it is the identity matrix at $Z=Z_{i-1}$.

To use~\eqref{WLtws1} in a correlation function, we will need to express the holomorphic frame $H(x_i,\lambda)H(x_i,\lambda_{i-1})^{-1}$ in terms of the twistor field $a$ that will appear in the action. This may be done as follows. Let $\delbar^{-1}_i$ be the inverse d-bar operator on the line X$_i$, with the boundary condition that it vanishes at the point $Z_{i-1}$.  Explicitly, if we parametrise the line X$_i$ by
\be
	Z(s)_i=s Z_{i-1}+Z_i\, , \qquad s\in \C.
\ee
then for any $(0,1)$-form $\omega$ on X$_i$, then the function $\delbar^{-1}_i\omega$ may be defined by 
\be
	(\bar\del^{-1}_i \omega)(s)\equiv\int_{{\rm X}_i} G(s,s')\wedge \omega(s')\, ,
\ee
where the Greens function
\be	
	G(s,s')= \frac{1}{2\pi\im}\frac{\rd s'}{(s-s')}
\ee
(and hence $\bar\del^{-1}_i\omega$) vanishes at $Z_{i-1}$ (where $s=\infty$). Solving~\eqref{Hdef} perturbatively shows that  $H(x_i,\lambda) H(x_{i},\lambda_{i-1})^{-1}$ is given by the infinite series 
\be
	H(x_i,\lambda) H(x_{i},\lambda_{i-1})^{-1}= \sum_{l=0}^\infty (\bar\del_i^{-1}a(Z(s)_i))^l 
	= \left.\delbar_{{\rm X}_i}\frac{1}{\delbar+a}\right|_{{\rm X}_i}\,,
\ee
where again we have to take care that the terms in $(\bar\del_i^{-1} a)^l $ are appropriately ordered.  Thus we obtain the perturbative expansion
\be
	{\rm W}[C_n] = \tr \, {\rm P} \prod_{i=1}^n \left( \sum_{l_i=0}^\infty\, (\bar\del_i^{-1}a(Z(s)_i))^{l_i} \right)
\label{WLtws2}
\ee
for the null polygonal Wilson loop in terms of the twistor field $a$. Exactly as on space-time, computing the correlation function of this operator using the twistor action for $\cN=4$ SYM gives the $n$-particle MHV amplitudes, as we shall see below.


\subsection{A supersymmetric Wilson Loop}
\label{sec:susyWL}

$\cN=4$ SYM naturally lives on supertwistor space $\CP^{3|4}$, rather than on $\CP^3$. Thus, instead of a connection (0,1)-form $a(Z)$ we really have an $\cN=4$ superfield
\be
	\cA(Z,\chi) = a(Z) + \chi_a\,\gamma^a(Z)+\frac{1}{2}\chi_a\chi_b\,\phi^{ab}(Z) 
	+ \frac{\epsilon^{abcd}}{3!}\chi_a\chi_b\chi_c\,\tilde\gamma_d 
	+ \frac{\epsilon^{abcd}}{4!}\chi_a\chi_b\chi_c\chi_d\, g(Z)
\label{superfield}
\ee
that may be thought of as a connection (0,1)-form on a bundle over $\CP^{3|4}$. The lowest component of $\cA$ is the non-supersymmetric field $a$, while the Grassmann expansion of $\cA(Z,\chi)$ provides the rest of the supermultiplet. As we will see in the next section, the twistor action for $\cN=4$ SYM can be written in terms of the superfield $\cA$, fully off-shell.

From this perspective, it is unnatural to define the Wilson loop purely in terms of the non-supersymmetric field $a$ as in~\eqref{WLtws2}. Instead, we replace $a$ by the superfield $\cA$ to find an $\cN=4$ supersymmetric Wilson loop on twistor space (that we also call ${\rm W}[C_n]$):
\be
	{\rm W}[C_n] \equiv \tr \, {\rm P} \prod_{i=1}^n \left( \sum_{l_i=0}^\infty\, (\bar\del_i^{-1}\cA(Z(s)_i))^{l_i} \right)\, .
\label{WLtws}
\ee	
It is this operator that we use in the rest of the paper. The correlation function of this supersymmetric Wilson loop will provide us with the complete planar S-matrix of the $\cN=4$ theory, including all N$^k$MHV amplitudes.  We return to the definition of this Wilson-loop in space-time in section~\ref{sec:space-time}.


\section{The Twistor Action for $\cN=4$ SYM}
\label{sec:action}

$\cN=4$ super Yang-Mills may be defined on twistor space by the action~\cite{Boels:2006ir} 
\be
	S=S_1+S_2\,,
\label{action}
\ee
where $S_1$ is a holomorphic Chern-Simons action~\cite{Witten:1992fb,Witten:2003nn}
\be
	S_1[\cA] 
	= \int_{\CP^{3|4}}\hspace{-0.3cm}{\rm D}^{3|4}Z\wedge {\rm Tr}\left(\!\cA\,\delbar\cA+\frac{2}{3}\cA^3\!\right) 
\label{action1}
\ee
for the connection (0,1)-form $\cA$ on $\CP^{3|4}$, and $S_2$ is a non-local term
\be
	S_2[\cA] = {\rm g}^2\int_\Gamma {\rm d}^{4|8}x\  \ln\det\left.\Dbar\right|_{\rm X}
\label{action2}
\ee
consisting of the integral of the logarithm of the determinant\footnote{
		This determinant can  be understood either via the Quillen construction~\cite{QuillenLine}, or,
		as in twistor-string theory~\cite{Witten:2003nn,Mason:2005zm, Boels:2006ir},  as the partition
		function of a chiral free fermion CFT on X. Technically, the $\Dbar$-operator in this operator must
		be taken to act on sections of $E|_{\rm X}\otimes \cO_{\rm X}(-1)$ so that it has vanishing index and
		the determinant exists. Physically, the twisting accounts for the spins of the fermions on X, and ensures
		the CFT has no axial anomaly.
		}
of the d-bar operator $\Dbar=\bar\del +\cA$, restricted to a line ${\rm X}\subset\CP^{3|4}$.  This object is then integrated over a contour $\Gamma$ in the space of lines, corresponding to a real slice of compactified, complexified space-time. Thus the action is defined on the space of connection (0,1)-forms $\cA$ on a complex bundle $E\to\CP^{3|4}$ that is trivial upon restriction to every line in $\Gamma$. See \cite{Boels:2006ir} for further discussion.

Taken alone, the field equations of $S_1$ state that $\Dbar^2=F^{0,2}=0$, so that $E$ becomes a holomorphic bundle. The Penrose-Ward transform~\cite{Ward:1977ta} states that such holomorphic bundles on twistor space are in one-to-one correspondence with self-dual Yang-Mills bundles on space-time.  The supersymmetric extension provides the rest of the multiplet (including the anti self-dual part of the Yang-Mills field) as  linearized fields coupled to this self-dual background. The role of $S_2$ is to promote the action to describe full, rather than anti self-dual, $\cN=4$ SYM (at least perturbatively). This can be seen by expanding $\ln\det\Dbar|_{\rm X}$ as the infinite series
\be
	\ln\det\left.\Dbar\right|_{\rm X} 
	= {\rm Tr}(\left.\!\ln\delbar\right|_{\rm X}) + \sum_{m=2}^\infty\frac{1}{m}\int_{{\rm X}^m}\hspace{-0.1cm}
	{\rm Tr}\left(\delbar^{-1}\!\cA_1\,\delbar^{-1}\!\cA_2\,\cdots\,\delbar^{-1}\!\cA_m\right)\,,
\label{expansion}
\ee
where $\cA_i$ is the field $\cA$ inserted at some point $p_i\in{\rm X}$. These insertions are connected together using the free fermion propagator, or equivalently, the inverse Cauchy-Riemann operator $\delbar^{-1}$ acting on sections of $\cO_{\rm X}(-1)$.  Each term\footnote{The term with $m=1$ is absent because of the colour trace.} in this expansion is an $m$-particle MHV vertex, while the holomorphic Chern-Simons term contains the remaining 3 particle $\MHVbar$ vertex.

Both the action and the Wilson loop~\eqref{WLtws} are invariant\footnote{Twistor space has vanishing third Betti number, so gauge invariance of the holomorphic Chern-Simons term is automatic. Because $\det(\Dbar)$ is a section of a Quillen determinant line bundle, it is well-defined only up to an overall (complex) phase, so $\ln\det(\Dbar)$ is defined only up to an additive piece. This term is annihilated by the fermionic integration in ${\rm d}^{4|8}x$; see~\cite{Boels:2006ir}.} under general complex gauge transformations
\be
	\Dbar\ \mapsto \ g^{-1}\Dbar g\,,
\ee
where $g(Z,\bar Z)$ is a smooth map from twistor space to the gauge group. Twistor space has six (bosonic real) dimensions, so this is a considerably greater gauge freedom than in space-time.  As explained in~\cite{Boels:2006ir}, this gauge freedom may be exploited to relate the twistor action either to the MHV diagram formalism or to the standard space-time action for $\cN=4$ SYM. In the next section, we pick an axial gauge in which it is straightforward to evaluate the correlation function~\eqref{main}, obtaining a correspondence with the MHV diagram formalism. On the other hand, one can reduce to space-time by imposing the partial gauge fixing that $\cA|_{{\rm X}}$ is harmonic with respect to an arbitrary Hermitian metric on the Riemann sphere. In this gauge, the first two component fields in~\eqref{superfield} vanish upon restriction to X so that $\cA|_{\rm X}$ is of order $\chi^2$ and the expansion~\eqref{expansion} terminates. $S_2$ then reduces to
\be
	S'_2={\rm g}^2\int{\rm d}^4x\ {\rm Tr}\left\{2G_{AB}G^{AB}+\Phi^{ab}\,\Psi_{Aa}\Psi^{A}_{\ b} 
	+ \frac{1}{4}\Phi^{ab}\Phi_{bc}\Phi^{cd}\Phi_{da}\right\}
\label{S2spacetime}
\ee
on space-time, where $\Phi$ is a space-time scalar field, $\Psi_A$ a left Weyl spinor and $G_{AB}\,{\rm d}x^{AA'}\wedge{\rm d}x^B_{\ A'}$ an anti self-dual 2-form corresponding to the twistor fields $\phi(Z)$, $\tilde\gamma(Z)$ and $g(Z)$, respectively (see~\cite{Boels:2006ir,Lovelace:2010ev} for details). Likewise, in this gauge $S_1$ may be reduced to the space-time form
\be
	S'_1 = \int{\rm d}^4x\ {\rm Tr}\left\{ G^{AB}F_{AB} + \widetilde\Psi^{A'a}D_{AA'}\Psi^A_{\ a} 
	+ \frac{1}{2} D_{AA'}\Phi^{ab}\,D^{AA'}\Phi_{ab} + \Phi_{ab}\widetilde\Psi^{A'a}\widetilde\Psi_{A'}^b\right\}
\ee
and $S'_1+S'_2$ is the Chalmers-Siegel action for perturbative $\cN=4$ SYM~\cite{Siegel:1992za,Chalmers:1996rq}.

Because the twistor action is fully off-shell,  its associated path integral may be used to compute arbitrary correlation functions as well as scattering amplitudes. In this paper we focus on the correlator~\eqref{main}, but the techniques have much wider applicability, as will be explored in~\cite{withMat}. The proposal that the this correlation function reproduces the usual scattering amplitudes (divided by the MHV tree amplitude) arises if we declare that the action~\eqref{action} and correlator both live in \emph{momentum} twistor space~\cite{Hodges:2009hk,Mason:2009qx}. This is completely analogous to computing the expectation value of a polygonal Wilson loop using the standard, space-time $\cN=4$ SYM action and then comparing to MHV amplitudes by declaring the computation to have taken place on region momentum space (\emph{dual} space-time).


\subsection{Correlation functions from the twistor action in an axial gauge}
\label{sec:examples}

In this section, we perturbatively compute the correlation function~\eqref{main} using the twistor action in the axial gauge  
\be
	\bar Z_\CSW^{\bar\alpha}\frac{\del}{\del \bar Z^{\bar\alpha}}\lrcorner\,\cA =0\, ,
\label{axial}
\ee
where $Z_*$ is a fixed reference twistor. This gauge greatly simplifies the evaluation of the correlation function, because the cubic Chern-Simons vertex is eliminated leaving just the MHV vertices in $S_2$. However, since these vertices all come with a factor of the Yang-Mills coupling g$^2$, they are irrelevant when computing the correlator at lowest order (g$^0$). Thus, at order g$^0$, the correlator is evaluated
\be
	\int [{\rm d}\cA] \ {\rm W}[C_n]\, \exp(-S_1[\cA])
\ee
in holomorphic Chern-Simons theory.   by pairwise contracting fields in W$[C_n]$ with axial gauge propagators $\Delta_*$ of the holomorphic Chern-Simons theory.   According to section~\ref{sec:WL} we can expand the operator as
\be
	{\rm W}[C_n] = \tr \, {\rm P} \prod_{i=1}^n \left( \sum_{l_i=0}^\infty\, (\delbar_i^{-1}\cA_i)^{l_i} \right)
\label{expansion2}
\ee
where the appropriate cyclic ordering in the matrix products is understood.  In the correlation function, only terms with an even number of $\cA$ insertions contribute at this order. We will show that the correlator of the $m=2k$ term in~\eqref{expansion2} computes the ratio of the N$^k$MHV tree amplitude to the MHV tree.

At higher orders in the coupling we can also make use of the MHV vertices~\eqref{expansion} in $S_2$. Each such vertex comes with an integral over the choice of line ${\rm X}\subset\CP^{3|4}$, or equivalently an integral over (dual conformal) space-time. From the point of view of scattering amplitudes, these integrals will be the usual loop integrals. The expansion of $\la {\rm W}[C_n]\ra$ in powers of g$^2$ is thus the loop expansion of the amplitude, with the number of loops equal to the number of MHV vertices used in the correlator. Note that $S'_2$ -- the space-time equivalent~\eqref{S2spacetime} of the sum of all MHV vertices -- is closely related to the chiral Lagrangian insertions used by~\cite{Eden:2010ce} to obtain the integrand of certain loop amplitudes, here in an $\cN=4$ context.

In this paper, we do not attempt to evaluate the loop integrals themselves -- they are divergent and require regularisation. However,  these integrals can be regulated and evaluated directly in twistor space. This has been done in~\cite{Hodges:2010kq,Mason:2010pg} at one loop and in~\cite{Drummond:2010mb,Alday:2010jz} for certain two loop integrals, using the Coulomb branch (or AdS) regularisation scheme introduced in~\cite{Alday:2009zm,Henn:2010bk}. Higher-loop MHV diagrams have also been considered in~\cite{Bern:2009xq,Sever:2009aa}.

In this axial gauge, the Feynman diagrams of the twistor action are  MHV diagrams~\cite{Cachazo:2004kj, Boels:2007qn}, with the CSW reference spinor determined by the fixed reference twistor $Z_\CSW$ (the reference twistor is usually taken at infinity, where its information is that of a spinor). We will see that, remarkably, the space-time MHV diagrams generated by the correlator~\eqref{main} are the planar duals of the usual MHV diagrams for the corresponding scattering amplitude.


\subsection{The propagator in an axial gauge}
\label{sec:propagator}

The first ingredient we need is the propagator of the holomorphic Chern-Simons theory. In axial gauge this takes the particularly simple form
\be
	\Delta_\CSW(Z,Z') = \bar\delta^{2|4}(Z,Z',Z_\CSW) \equiv 
	\int_{\CP^2}\frac{{\rm D}^2c}{c_1c_2c_3}\wedge\bar\delta^{4|4}(c_1Z+c_2Z'+c_3Z_\CSW)
\label{propagator}
\ee
where $D^2c=c_1\rd c_2\wedge \rd c_3 +$ cyclic.  This delta function restricts $Z$ and $Z'$ to be collinear with the reference twistor $Z_\CSW$ (in the projective space; see~\cite{Mason:2009sa, Mason:2009qx} for further discussion of such projective $\bar\delta$-functions). To understand this definition, recall~\cite{Cachazo:2004kj,Atiyah:1981ey} that the propagator for the $\cA$ field is the inverse of the $\delbar$-operator on $\CP^{3|4}$, so may be represented by a (0,2)-form $\Delta$ on $\CP^{3|4}\times\CP^{3|4}$ that has homogeneity zero in each entry and satisfies
\be
	\delbar_Z \Delta(Z,Z') = \delbar_{Z'} \Delta(Z,Z') = \bar\delta^{3|4}(Z,Z') \,,
\label{dbar-prop}
\ee	
where the $\bar\delta$-function on the right is a $(0,3)$-form supported on the diagonal $\CP^{3|4}\subset\CP^{3|4}_Z\times\CP^{3|4}_{Z'}$.  The only possible singularities of $\bar\delta^{2|4}(Z,Z',Z_\CSW)$ occur when any two of the three points collide, where being collinear is no longer a restriction. Indeed, a direct calculation using the definition~\eqref{propagator} gives
\be
	\delbar \Delta_\CSW(Z,Z')=\bar\delta^{3|4}(Z,Z') + \bar\delta^{3|4}(Z',Z_\CSW)+\bar\delta^{3|4}(Z_\CSW,Z) \, .
\ee
The first term is the required singularity of the propagator on the diagonal. The remaining two terms are `spurious' singularities that are a common feature of axial gauges. A proper study of this propagator, including its relation to more standard momentum space propagators and its use in the MHV diagram formalism for scattering amplitudes on ordinary twistor space, will be presented in~\cite{Adamo:2010}.


\section{Tree amplitudes}
\label{sec:trees}

In this section we compute the NMHV and N$^2$MHV tree amplitudes from the Wilson loop.  The answers are given as a sum of products of dual superconformal invariants, and agree with the results obtained from the momentum twistor reformulation of the MHV rules derived in~\cite{CSWMat}. There it was shown that the MHV diagram rules in momentum twistor space associated just 1 to each vertex, and a dual-conformal R-invariant \eqref{Rinv} to each propagator.  Here we will similarly see an R-invariant arising for each propagator but this will be associated to the planar dual of the MHV diagram.


\subsection{NMHV}

The zeroth order term is just $\la \tr\, ({\rm Id}) \ra$.  For an SU$(N)$ gauge group, we divide by $N$ to normalize this term to 1.  The first nontrivial term to compute is thus
\be
	\sum_{1\leq j\leq i \leq n}\big\la\tr \,(\delbar_i^{-1}\!\cA_i\,\delbar_j^{-1}\!\cA_j\big\ra_{\cO({\rm g}^0)}\,,
\ee
where each field is inserted at some point on the components X$_i$ and X$_j$ of the curve $C_n$. (The integration over 
X$_i$ and X$_j$ is implicit in the definition of $\bar\del^{-1}$.)

Suppose first that $i\neq j$ so that the fields are inserted on different components (see figure~\ref{fig:NMHVcorr}). We may parametrize their locations by
\be
	Z(s) = sZ_{i-1}+Z_i\qquad\hbox{and}\qquad Z(t)=tZ_{j-1}+Z_j
\ee
in terms of local coordinates $s$ and $t$ on the two lines. The operator $\delbar^{-1}|_{{\rm X}_i}$  was defined as the Green function on X$_i$ that vanishes at $Z_{i-1}$,  {\it i.e.}, at $s=\infty$.  Thus the Greens functions are simply ${\rm d}s/s$ and ${\rm d}t/t$. Contracting the two fields with a single twistor space propagator gives
\be
\begin{aligned}
	\la\tr \,(\delbar_{ji}^{-1}\!\cA_i\,\delbar_{ij}^{-1}\!\cA_j\big\ra
	&= \int\frac{{\rm d}s}{s}\frac{{\rm d}t}{t}\,\Delta_\CSW(Z(s),Z(t))\\
	&=\int\frac{{\rm d}^4t}{t_1t_2t_3t_4} \bar\delta^{4|4}(Z_\CSW+t_1Z_{i-1}+t_2Z_i+t_3Z_{j-1}+t_4Z_j)
\end{aligned}
\label{basicintegral}
\ee
where the second line follows from the definition~\eqref{propagator} of the propagator, after a rescaling of the integration variables. Without further calculation, we can recognise~\eqref{basicintegral} as a dual superconformal `R-invariant' in the momentum twistor form\footnote{In the notation of~\cite{Drummond:2008vq}, $[\,\CSW\,,i\!-\!1,i,j\!-\!1,j] = R_{\CSW;ij}$. We use the notation $[\ ,\ ,\ ,\ ,\ ]$ to emphasise that the R-invariant is a totally antisymmetric function of five supertwistors, homogeneous of degree zero in each entry. We also define $\la a,b,c,d\ra$ to be the skew product of the bosonic components of four supertwistors $Z_a,\ldots,Z_d$. This notation was introduced in~\cite{Arkani-Hamed:2010kv}.}
\be
\begin{aligned}
	{[a,b,c,d,e]} &\equiv \int\frac{{\rm d}^4t}{t_1t_2t_3t_4} \bar\delta^{4|4}(Z_a+t_1Z_b+t_2Z_c+t_3Z_d+t_4Z_e)\\
	&=\frac{\delta^{0|4}(\psi_a\la b,c,d,e\ra + \hbox{cyclic})}
	{\la a,b,c,d\ra\,\la b,c,d,e\ra\,\la c,d,e,a\ra\,\la d,e,a,b\ra\,\la e,a,b,c\ra}
\end{aligned}
\label{Rinv}
\ee
first introduced in~\cite{Mason:2009qx}. Thus we have immediately 
\be
	\la\tr \,(\delbar_i^{-1}\!\cA_i\,\delbar_j^{-1}\!\cA_j\big\ra
	= [\,*\,,i\!-\!1,i,j\!-\!1,j]
\ee
as the contribution to the correlator from two fields inserted on two different lines, as illustrated in figure~\ref{fig:NMHVcorr}.

\FIGURE[t]{
	\includegraphics[height=40mm]{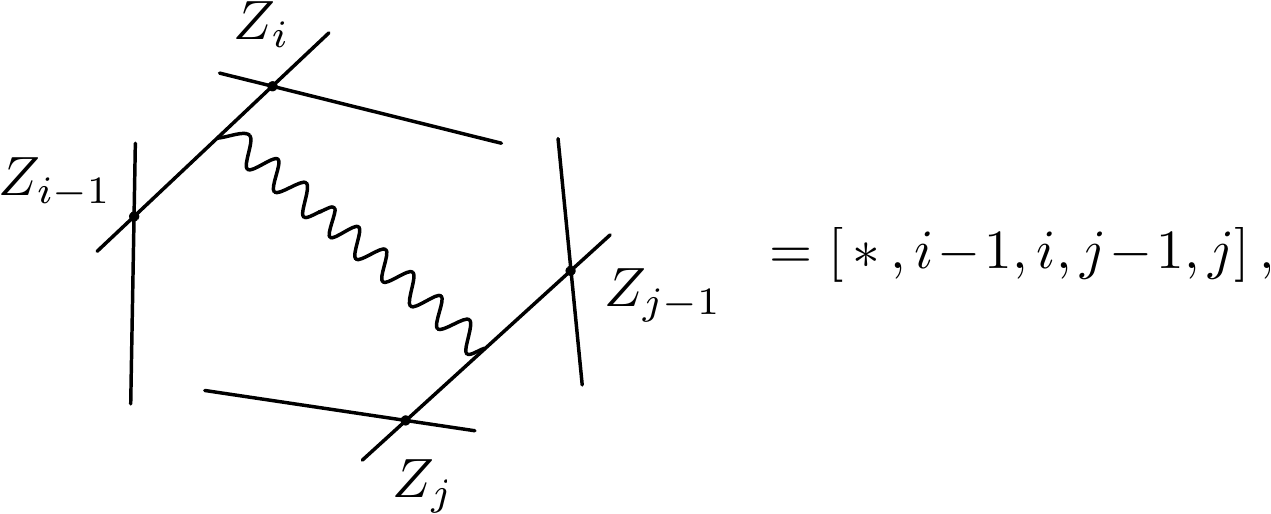}
	\caption{A single propagator connecting two $\cA$ insertions on lines X$_i$ and X$_j$ corresponds to the basic
			dual superconformal invariant $[\,*\,i\!-\!1,i,j\!-\!1,j]$. (The diagram shows only the relevant lines 
			of $C_n$.}
	\label{fig:NMHVcorr}
}

We now sum over the possible locations for the field insertion. The R-invariant vanishes when two or more of its arguments are the same because $[\,\CSW\,,i\!-\!1,i,j\!-\!1,j]$ is totally antisymmetric and homogeneous of degree zero.  Thus the contribution vanishes whenever the two fields are inserted either on the same line or on adjacent components of $C_n$. Summing over components therefore gives 
\be
	\left\la{\rm Tr}\,(\delbar^{-1}\!\cA\,\delbar^{-1}\!\cA)\right\ra
	= \ \sum_{1\leq i<j \leq n}[\,\CSW\,,i\!-\!1,i,j\!-\!1,j]
\label{NMHV0}
\ee
to order g$^0$, where the inverse d-bar operators here are understood to live on $C_n$ ({\it i.e.}, they include the sum over components of $C_n$ in their definition).  This is the precisely the momentum twistor expression
\be
	M_{\rm NMHV}^{(0)} = \sum_{i<j}\, [\,\CSW\,,i\!-\!1,i,j\!-\!1,j]\,.
\ee
for the $n$-particle NMHV tree amplitude (divided by the MHV tree) found in~\cite{CSWMat} using the MHV diagram formalism.  In terms of MHV diagrams for the scattering amplitude, each term in this sum comes from a diagram that connects two MHV vertices with a single propagator. Notice also that if we choose the reference twistor $Z_\CSW$ to equal one of the external momentum twistors, say $Z_n$, then~\eqref{NMHV0} reduces to the standard BCFW form of the NMHV tree~\cite{Drummond:2008vq}. 


\subsection{N$^2$MHV}

\begin{figure}[t]
	\centering
	\includegraphics[height=50mm]{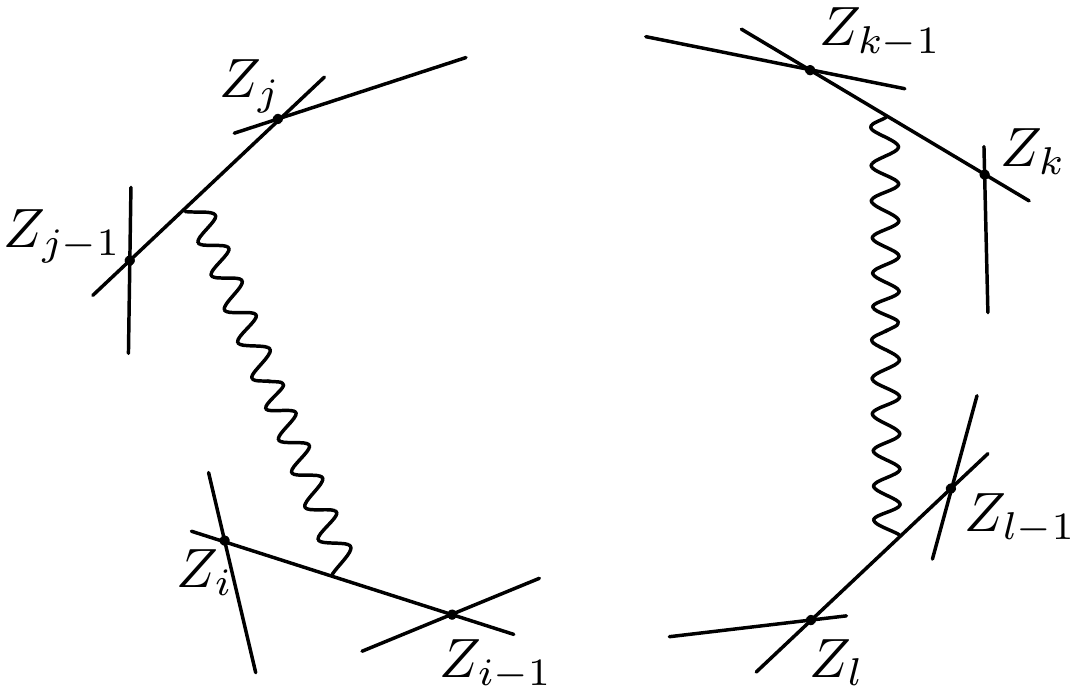}
	\caption{Correlation function diagrams with two propagators (and no vertices) contribute to the N$^2$MHV
			tree amplitude. Only those diagrams whose contractions are compatible with the ordering of the
			trace in the Wilson loop survive in the planar limit.} 
\label{fig:treeNNMHVtwistor}
\end{figure}

The next non-vanishing term at order g$^0$ contains four powers of $\cA$. Consider first the case where these fields are each inserted on different lines, say
\be
	Z(s) = sZ_{i-1}+Z_i, \quad Z(t)=tZ_{j-1}+Z_j,\quad Z(u)=uZ_{k-1}+Z_k, \quad Z(v) = vZ_{l-1}+Z_l, .
\ee
Although all possible contractions are allowed, only those in which the contracted $\cA$s are adjacent in the trace ordering survive in the planar limit, the alternating contraction being suppressed by a factor of $1/N$. Thus,  only the diagrams where the twistor propagators do not cross survive in the planar limit (see figure~\ref{fig:treeNNMHVtwistor}), although we can allow $i=l$ and/or $j=k$.

In the generic case $i\neq l$ and $j\neq k$ we find
\be
\begin{aligned}
	&\int\frac{{\rm d}s}{s}\frac{{\rm d}t}{t}\frac{{\rm d}u}{u}\frac{{\rm d}v}{v}\,\Delta_\CSW(Z(s),Z(t))\,\Delta_\CSW(Z(u),Z(v))\\
	&\hspace{4cm}= [\,\CSW\,,i\!-\!1,i,j\!-\!1,j]\,[\,\CSW\,,k\!-\!1,k,l\!-\!1,l]\,,
\label{NNMHVgeneric}
\end{aligned}
\ee
where each factor of the integration measure is provided by an inverse $\delbar$-operator on $C_n$.  Once again, the identification as a product of two dual superconformal invariants (depending on the axial gauge reference twistor $Z_\CSW$) follows immediately from the form~\eqref{propagator} for the propagator.

Finally, there is an exceptional class of diagrams to consider where more than one field is inserted on the same twistor line. Suppose first that $j=k$ so that we have
\be
	Z(s) = s Z_{i-1}+Z_i,\quad Z(t_1)=t_1Z_{j-1}+Z_j,\quad Z(t_2)=t_2Z_{k-1}+Z_k, \quad Z(v) = vZ_{i-1}+Z_i\,.
\ee
There is no difference in the form of the propagators $\Delta_\CSW(Z,Z')$, nor in the $\delbar^{-1}$-operators on the lines $(j\!-\!1,j)$ and $(k\!-\!1,k)$. However, compared to the measure in~\eqref{NNMHVgeneric} we must replace
\be
	\frac{{\rm d}t}{t}\frac{{\rm d}u}{u} \to \frac{{\rm d}t_1}{(t_2-t_1)}\frac{{\rm d}t_1}{t_1}
\ee
because, using the cyclicity of the trace and reading from the right, the $\delbar^{-1}$ operators in $\cdots\delbar^{-1}\cA(s_2)\,\delbar^{-1}\cA(s_1)\,\delbar^{-1}\cdots$ first link the node at $Z_i$ (where $s_i=\infty$) to $\cA(s_1)$, then link $\cA(s_1)$ to $\cA(s_2)$ on the same line component, and then link $\cA(s_2)$ to the node at $Z_{i-1}$ (where $s_i=0$). Consequently, this term gives a contribution
\be
\begin{aligned}
	&\int\frac{{\rm d}s}{s}\frac{{\rm d}t_1}{t_1}
\frac{{\rm d}t_2}{t_2-t_1}\frac{{\rm d}v}{v}\,
	\Delta_\CSW(Z(s_1),Z(t))\,\Delta_\CSW(Z(u),Z(s_2))\\
	&\hspace{5cm}= [\,\CSW\,,{i\!-\!1},i,j\!-\!1,j]\,[\,\CSW\,,\widehat{k\!-\!1},k,i\!-\!1,i]
\end{aligned}
\label{NNMHVboundary}
\ee
that differs from~\eqref{NNMHVgeneric} only by replacing $Z_{k-1}$ by the intersection of the line $(k\!-\!1,k)$ with the plane $(\,*\,,i\!-\!1,i)$:
\be
\begin{aligned}
	Z_{k-1}\,  \to\,  \widehat{Z_{k-1}} &\equiv (k\!-\!1,k)\cap(\,*\,,i\!-\!1,i)\\
					&=\la *,i\!-\!1,i,k\!-\!1\ra Z_k - \la *,i\!-\!1,i,k\ra Z_{k-1}\, .
\end{aligned}
\ee
This follows from a simple change of variables $t_2 \rightarrow w=t_2-t_1$ in the integrand of~\eqref{NNMHVboundary},  with $\widehat{Z_{k-1}}$  identified as the coefficient of $w$ inside the $\delta$-function.

Terms in which both propagators end on the same pair of lines, or where one of the propagators has both ends on the same line vanish identically, while terms in which the $\Delta_\CSW$ propagators connect the fields in a different order (so that $s_1$ and $s_2$ are exchanged in the twistor propagators of~\eqref{NNMHVboundary}, but not in the measure) are suppressed in the planar limit. Therefore, summing over the possible lines on which the propagators end gives
\be
	\left\la{\rm Tr}\left(\delbar^{-1}\!\cA\,\delbar^{-1}\!\cA\,\delbar^{-1}\!\cA\,\delbar^{-1}\!\cA\right)\right\ra
	= \hspace{-0.4cm} \sum_{1\leq i<j\leq k<l\leq n+i}\hspace{-0.5cm}
	 [\,\CSW\,,\widehat{i\!-\!1},i,j\!-\!1,j]\,[\,\CSW\,,\widehat{k\!-\!1},k,l\!-\!1,l]
\label{NMHV}
\ee
to order g$^0$, where the hatted variables are defined as
\be
	\widehat{i\!-\!1}=
		\begin{cases}
			(i\!-\!1,i)\cap(\,\CSW\,,k\!-\!1,k) & \hbox{if } l= i\\
			\,i\!-\!1 & \hbox{otherwise,} 
		\end{cases}
\label{shift1}
\ee
and
\be
	\widehat{k\!-\!1}=
		\begin{cases}
			(k\!-\!1,k)\cap(\,\CSW\,,i\!-\!1,i) &  \hbox{if } k=j\\
			\,k\!-\!1 & \hbox{otherwise.}
		\end{cases}
\label{shift2}
\ee
This is precisely the momentum twistor form of the N$^2$MHV tree amplitude, divided by the MHV tree, as found in~\cite{CSWMat}. From the perspective of MHV diagrams for the scattering amplitude, term by term, each summand in~\eqref{NMHV} comes from a diagram consisting of two propagators joining three MHV vertices. The shifts~\eqref{shift1}-\eqref{shift2} of the momentum twistors occur when two propagators border the same region, so that they are adjacent in the cyclic ordering of the middle MHV vertex (see figure~\ref{fig:dualtrees}).


\subsection{Dual MHV diagrams and N$^k$MHV amplitudes}
	
It is revealing to draw MHV diagrams corresponding to the twistor configurations for the correlation function as found above. Lines in twistor space correspond to points in space-time, so we can represent a twistor propagator whose ends are integrated\footnote{
		Given two arbitrary lines in twistor space, there is a unique line that passes through a generic point
		$Z_\CSW$ and intersects both given lines. Since $P_\CSW(Z,Z')$ requires $Z$, $Z'$ and
		$Z_\CSW$ to be collinear, it follows that for a generic choice of $Z_\CSW$, only one twistor
		propagator actually contributes to the integrals.
		}
over the lines $(i\!-\!1,i)$ and $(j\!-\!1,j)$ by a space-time propagator joining the two points $x_i$ and $x_j$. As shown in figure~\ref{fig:dualtrees}, the resulting diagrams for the correlation function are the planar duals of the usual MHV diagrams~\cite{Cachazo:2004kj} for the corresponding NMHV and N$^2$MHV tree amplitudes.

\FIGURE[t]{
	\includegraphics[height=25mm]{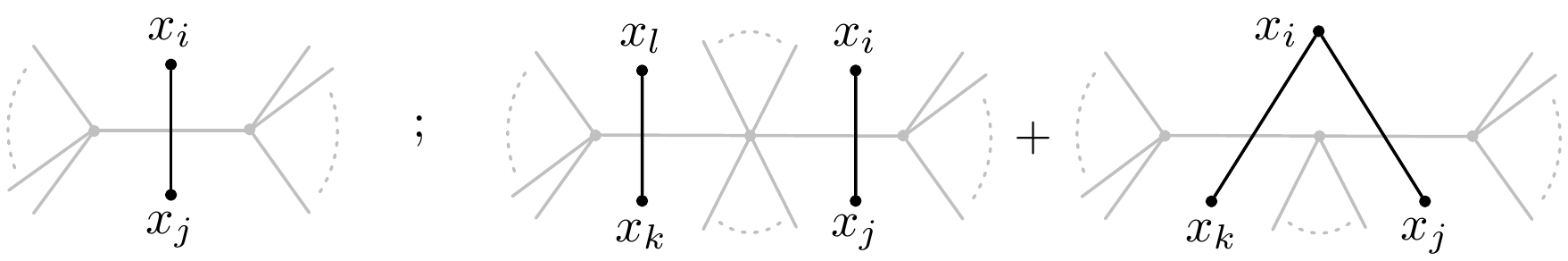}
	\caption{ The Feynman diagrams of the correlation function (shown in black) are dual to the MHV
	diagrams for the corresponding amplitude (shown in grey).}
	\label{fig:dualtrees}
}

Dualization leaves the number of propagators unchanged, so at order g$^0$, the N$^k$MHV tree amplitude will come from the expansion of $\la {\rm W}[C_n]\ra$ to order $2k$ in the field. These fields are contracted using $k$ propagators that, in the planar limit, do not cross each other.  The propagators therefore divide the polygon into $k+1$ regions, so the dual MHV diagram is a planar diagram with $k+1$ MHV vertices and $k$ propagators. As described in~\cite{CSWMat}, such MHV diagrams yield a product of $k$ R-invariants in momentum twistor space.  The boundary terms where two or more Wilson loop propagators end on the same edge correspond precisely to the boundary terms for the MHV diagrams in momentum twistor space, where two propagators are adjacent to the same region.  The shifts of the arguments of the R-invariants in~\cite{CSWMat} follow in exactly the same way as for the N$^2$MHV term above, and generalise to arbitrary N$^k$MHV amplitudes. It is therefore easy to see that our correspondence extends to arbitrary N$^k$MHV amplitudes.

It is also easy to see that, had we defined our twistor space Wilson loop using the \emph{non}-supersymmetric connection $\overline{D}=\delbar+a$ that corresponds to the standard space-time Wilson loop, then  all the correlation function diagrams discussed in this section would vanish. This is because the kinetic term of the holomorphic Chern-Simons theory on $\CP^{3|4}$ has a component expansion
\be
	\int{\rm D}^{3|4}Z \wedge\tr\,(\cA\,\delbar\cA) = \int{\rm D}^3Z\wedge\tr\,(g\,\delbar a + \cdots )
\ee
so the component propagator connects an $a$ field to a $g$ field, rather than connecting two $a$ fields. To zeroth order in the Yang-Mills coupling, there would be no way to contract any field insertions. With this non-supersymmetric operator, the only non-vanishing term would have been the trivial one, representing the MHV tree.


\section{Loop amplitudes}
\label{sec:loops}

At higher order in the Yang-Mills coupling, we may connect the field insertions in W$[C_n]$ via any of the MHV vertices in the action~\eqref{action2}. As explained in section~\ref{sec:action}, these vertices are supported on an arbitrary line ${\rm X}\subset\CP^{3|4}$, unrelated to the components of $C_n$. The choice of X is integrated out over a contour $\Gamma$ corresponding to a real slice of space-time. These (dual) space-time integrals, coming from the non-local term in the twistor action, are the loop integrals of the original amplitude.

We can rewrite the integral over space-time in a way that emphasises its superconformal invariance. If X is the line $(A,B)$, we have that
\be
	{\rm D}^{3|4}Z_A\wedge {\rm D}^{3|4}Z_B 
	={\rm d}^{4|8}x\wedge\frac{\la \lambda_A\,{\rm d}\lambda_A\ra\wedge\la\lambda_B\,{\rm d}\lambda_B\ra}			{\la\lambda_A\,\lambda_B\ra^2}\,,
\ee
where $\lambda_{A,B}$ define the location of $A$ and $B$ on $(A,B)$. The rest of the integrand is independent of these locations. Hence, if we modify the definition of the contour $\Gamma$ to include a factor of the anti-diagonal\footnote{That is, we pick a notion of complex conjugation $\lambda\to\bar\lambda$ and set $\lambda_B=\bar{\lambda}_A$. The $\lambda$-dependent measure $\la \lambda_A{\rm d}\lambda_A\ra\wedge\la\lambda_B{\rm d}\lambda_B\ra\,/\,\la\lambda_A\lambda_B\ra^2$ reduces to the standard K\"ahler form on this $S^2$.} $S^2\subset\CP^1_{\lambda_A}\times\CP^1_{\lambda_B}$ for each $x$, then the $\lambda_A,\,\lambda_B$ integrals simply yield 1 and the dependence on the specific reference points is removed. This choice of contour is particularly natural in twistor space. Although the new contour looks like $\mathbb{R}^4\times S^2$ locally, globally the $S^2$s can be chosen to  fibre over the space-time base in such way that the whole contour is the anti-diagonal $\CP^3\subset\CP^3_A\times\CP^3_B$ (though the integrand requires regularisation on this contour). The implied real slice of compactified, complexified space-time has topology $S^4$ and may be thought of as a rotation of the Euclidean real slice towards Lorentzian signature space-time, in accordance with the Feynman $\im\epsilon$-prescription. (See~\cite{Hodges:2010kq,Mason:2010pg} for further details.)


\subsection{MHV}

The action contains a two-point MHV vertex.  This  plays no role in the usual MHV diagram formalism (even as a vertex) because momentum conservation makes its numerator factor vanish.  However, the momentum conserving $\delta$-function is absent in the correlation function calculation.   Consequently, the first case to consider is
\be
	\left\la{\rm Tr}\left(\delbar^{-1}\!\cA\,\delbar^{-1}\cA\right)\right\ra
\ee
which gave the NMHV tree at order g$^0$. At order g$^2$, the two fields are each connected to the intermediate line $(A,B)$ -- the MHV vertex -- using $\Delta_\CSW$ propagators. If the fields are inserted on $C_n$ at
\be
	Z(t) = sZ_{i-1}+Z_i\qquad\hbox{and}\qquad Z(t)=tZ_{j-1}+Z_j
\ee
and the propagators meet the auxiliary line $(A,B)$ at
\be
	Z(u_1) = Z_A+u_1Z_B\qquad\hbox{and}\qquad Z(u_2)=Z_A+u_2Z_B\,,
\ee
then the contribution to the correlator is
\be
	\int_\Gamma{\rm D}^{3|4}Z_A\wedge{\rm D}^{3|4}Z_B
	\int \frac{{\rm d}s}{s}\frac{{\rm d}t}{t}\frac{{\rm d}u_1\,{\rm d}u_2}{(u_1-u_2)^2}
	\ \Delta_\CSW(Z(s),Z(u_1))\,\Delta_\CSW(Z(t),Z(u_2))\,.
\ee
The measure 
\be
	\frac{{\rm d}u_1\,{\rm d}u_2}{(u_1-u_2)^2}
\label{ABmeasure}
\ee
comes from the $\delbar^{-1}$ operators in the expansion~\eqref{expansion} of the MHV vertices in the action. It may be understood as a two-term current correlator in the fermionic CFT that generates the MHV vertices. 

\FIGURE[t]{
	\includegraphics[width=150mm]{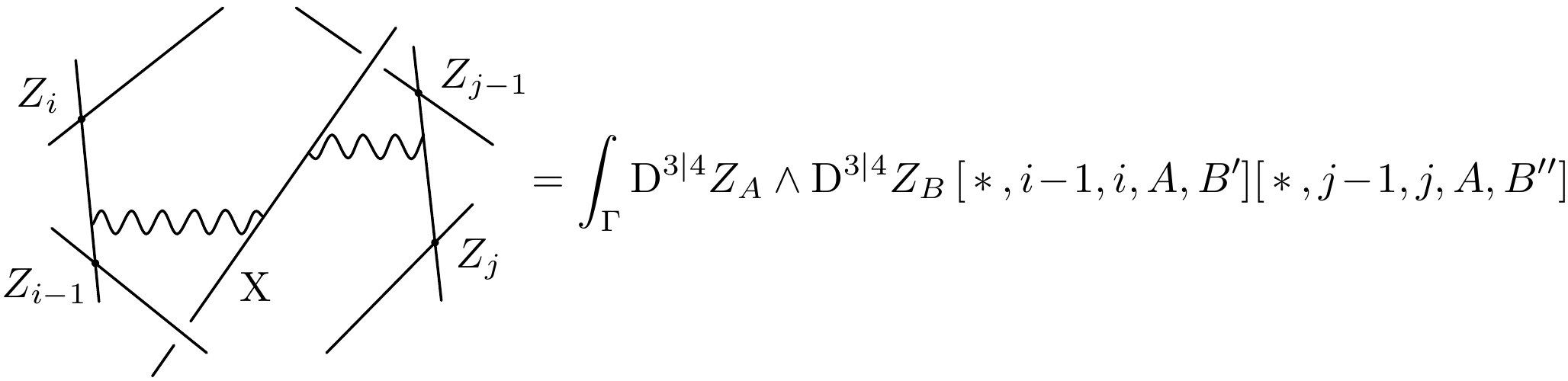}
	\caption{A particular configuration contributing to the 1-loop MHV amplitude.}
	\label{fig:1loopMHVtwistor}
}

It is again straightforward to identify the integrated propagators with dual superconformal invariants, with the result
\be
	\int_\Gamma{\rm D}^{3|4}Z_A\wedge{\rm D}^{3|4}Z_B\ [\,*\,,i\!-\!1,i,A,B']\,[\,*\,,j\!-\!1,j,A,B'']
\label{MHV1loopcontrib}
\ee
as shown in figure~\ref{fig:1loopMHVtwistor}. Just as in the boundary cases of the N$^2$MHV tree, the measure~\eqref{ABmeasure} mixes together the twistors $Z_A$ and $Z_B$ in the arguments of the $\delta$-functions. We can again choose a change of variables so that $B$ is shifted as
\be
	B' = (A,B)\cap(\,*\,,j\!-\!1,j) \qquad\hbox{and} \qquad B'' = (A,B)\cap (\,*\,,i\!-\!1,i)\,.
\ee
One can check that the two $\cA$ fields must be inserted on different components of $C_n$ for a non-vanishing contribution (although adjacent components do now contribute), so summing over the possible insertion points gives
\be
	\left\la{\rm Tr}\left(\delbar^{-1}\!\cA\,\delbar^{-1}\!\cA\,\right)\right\ra
	=\int_\Gamma{\rm D}^{3|4}Z_A\wedge{\rm D}^{3|4}Z_B\ \sum_{i<j}\, [\,\CSW\,,i\!-\!1,i,A,B']\,[\,\CSW\,,j\!-\!1,j,A,B'']
\label{MHV1}
\ee
at order g$^2$.

The integrand in~\eqref{MHV1} agrees identically with the integrand of the momentum twistor form of the $n$-particle 1-loop MHV amplitude as presented in~\cite{CSWMat}, which was proved to be equivalent to the expression for the 1-loop MHV amplitude obtained from the standard MHV rules in momentum space~\cite{Cachazo:2004kj,Brandhuber:2004yw,Bena:2004xu}. We thus have
\be
	\left\la{\rm Tr}\left(\delbar^{-1}\!\cA\,\delbar^{-1}\!\cA\,\right)\right\ra_{\cO({\rm g}^2)} = M_{\rm MHV}^{(1)}\, .
\ee
Once again, the summands in~\eqref{MHV1} correspond term by term to 1-loop MHV diagrams for the scattering amplitude. The arbitrary twistor line X corresponds to an arbitrary space-time point $x$, so the space-time diagram corresponding to figure~\ref{fig:1loopMHVtwistor} has a propagator connecting $x_i$ to $x$ and a further propagator connecting $x$ to $x_j$. As shown in figure~\ref{fig:1loopMHVST}, this diagram is nothing but the planar dual of an MHV diagram for the scattering amplitude.

\FIGURE[t]{
	\includegraphics[width=100mm]{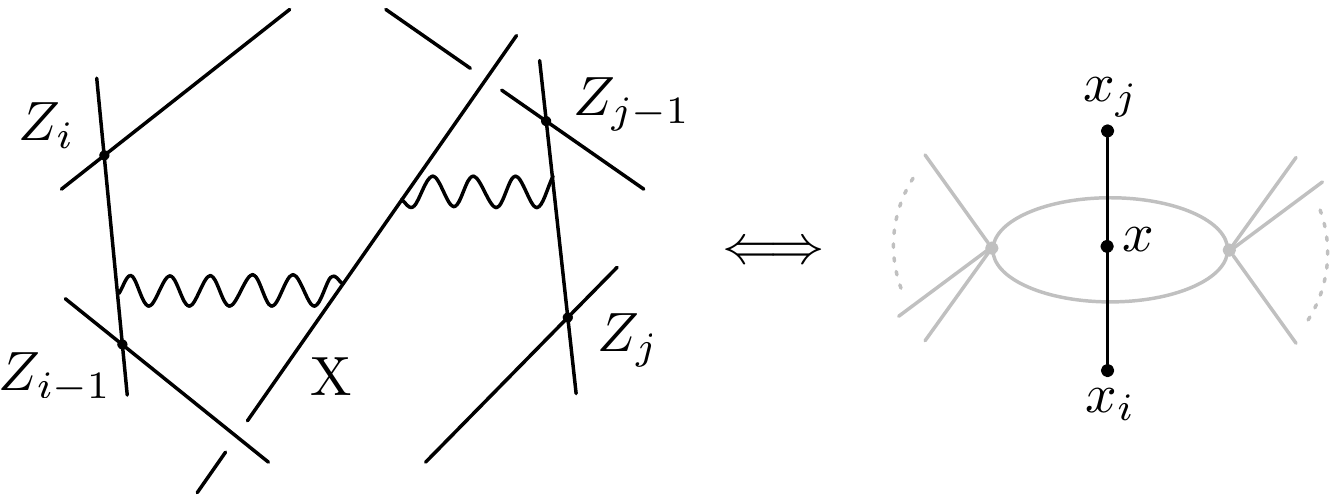}
	\caption{The space-time diagram for the correlation function is dual to the MHV diagram of the scattering amplitude.}
	\label{fig:1loopMHVST}
}


\subsection{NMHV}

\FIGURE[t]{
	\includegraphics[width=150mm]{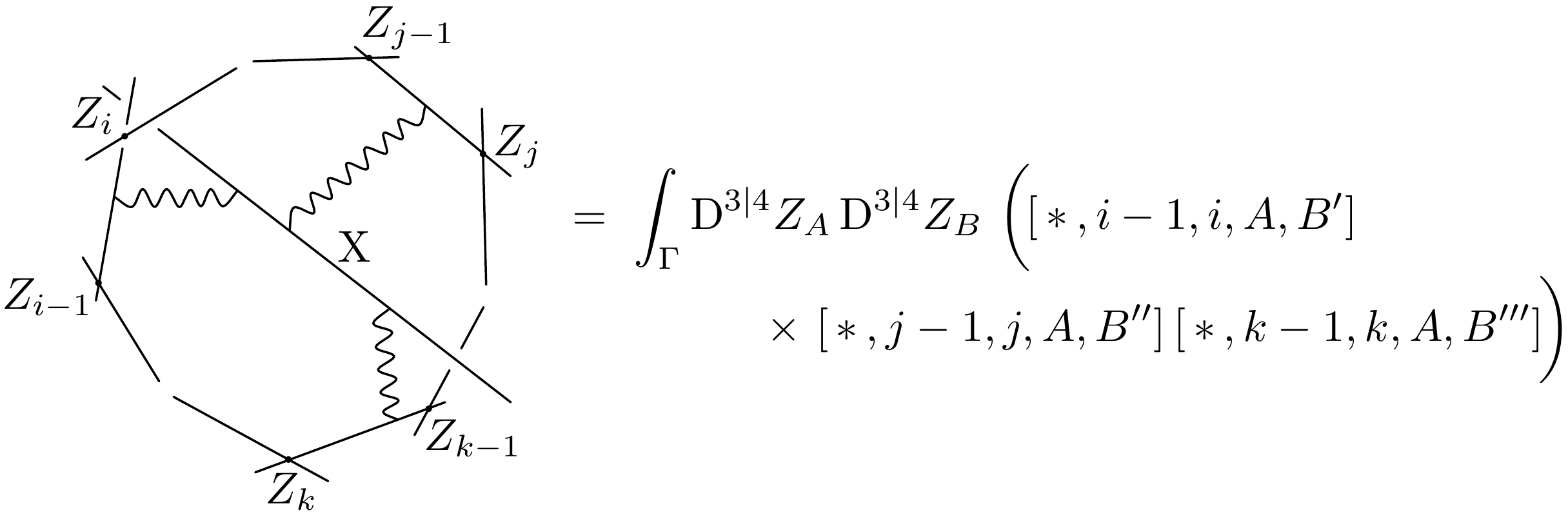}
	\caption{A contribution to the 1-loop NMHV amplitude coming from a cubic term in the field expansion of the 
			Wilson loop.}
	\label{fig:NMHVloopboundary}
}

As a further example, we consider the integrand of the planar 1-loop NMHV amplitude. Somewhat surprisingly, this amplitude receives contributions from two separate terms in the correlator; specifically, we shall show that $M_{\rm NMHV}^{(1)}$ comes from the order g$^2$ contribution to
\be
	\big\la{\rm Tr}\left(\delbar^{-1}\!\cA\,\delbar^{-1}\!\cA\,\delbar^{-1}\!\cA\,\right)\!\big\ra
	+\big\la{\rm Tr}\left(\delbar^{-1}\!\cA\,\delbar^{-1}\!\cA\,\delbar^{-1}\!\cA\,\delbar^{-1}\!\cA\,\right)\!\big\ra\,.
\label{1loopNMHVcorr}
\ee
The first term has three field insertions. When these are each on separate lines we obtain
\be
	\int_\Gamma{\rm D}^{3|4}Z_A\wedge{\rm D}^{3|4}Z_B\ 
	\frac{{\rm d} u_1\,{\rm d}u_2\,{\rm d}u_3}{(u_1-u_2)(u_2-u_3)(u_3-u_1)}\, 
	\prod_{a=1}^3\frac{{\rm d}s_a}{s_a}\,  \Delta_\CSW(Z(s_a),Z(u_a))
\label{NMHVloop1}
\ee
where
\be
	Z(s_1) = s_1Z_{i-1}+ Z_i,\qquad Z(s_2)= s_2Z_{j-1}+Z_j,\qquad Z(s_3) = s_3Z_{k-1}+Z_k
\ee
are the locations of the external fields, connected by propagators to points $Z(u_a) = Z_A+u_aZ_B$, for $a=1,2,3$ respectively. It is clear from the integrand that this term is a product of three dual superconformal invariants, each depending on the reference twistor $Z_\CSW$ and a pair of external twistors, together with some combination of the auxiliary twistors $Z_A$ and $Z_B$.  Performing a change of variables and identifying the R invariants, we find the contribution
\be
	\int_\Gamma{\rm D}^{3|4}Z_A\wedge{\rm D}^{3|4}Z_B\, [\,*\,,i\!-\!1,i,A,B']\,[\,*\,j\!-\!1,j,A,B'']\,[\,*\,,k\!-\!1,k,A,B''']
\label{NMHV1loopcontrib}
\ee
shown in figure~\ref{fig:NMHVloopboundary}. The shifts in loop variables again arise from changing variables in the integrand of~\eqref{NMHVloop1} and are defined as
\be
	B' = (AB)\cap(*,k\!-\!1,k)\,,\qquad B'' = (AB)\cap (*,i\!-\!1,i)\,, \qquad B''' = (AB)\cap(*,j\!-\!1,j)\, .
\ee
As at MHV, the corresponding diagram where two (or more) propagators end on the same component of $C_n$ can be shown to vanish, so in total we find
\be
\begin{aligned}
	&\left\la{\rm Tr}\left(\left.\delbar^{-1}\!\cA\,\delbar^{-1}\!\cA\,\delbar^{-1}\!\cA\right|_{C_n}\right)\right\ra\ = \\
	&\hspace{0.5cm}\int_\Gamma{\rm D}^{3|4}Z_A\wedge{\rm D}^{3|4}Z_B 
	\hspace{-0.2cm}\sum_{1\leq i<j<k\leq n}\hspace{-0.2cm}
	[\,\CSW\,,i\!-\!1,i,A,B']\,[\,\CSW\,j\!-\!1,j,A,B'']\,[\,\CSW\,,k\!-\!1,k,A,B''']
\label{NMHVloopboundary}
\end{aligned}
\ee
to order g$^2$.

\medskip

Diagrams that contribute to $\left\la{\rm Tr}\left(\delbar^{-1}\!\cA\,\delbar^{-1}\!\cA\,\delbar^{-1}\!\cA\,\delbar^{-1}\!\cA\right)\right\ra$ at order g$^2$ have the same form as the 1-loop MHV diagram, but with an extra twistor space propagator connecting the remaining two fields directly ({\it i.e.}, without connecting to $(A,B)$). If the external fields are inserted on distinct $C_n$ components
\be
	Z(s_1) = s_1Z_{i-1}+Z_i \quad Z(s_2)=s_2Z_{j-1}+Z_j\quad 
	Z(s_3)=s_3Z_{k-1}+Z_k \quad Z(s_4) = s_4Z_{l-1}+Z_l
\ee
as for the N$^2$MHV tree, then (suppressing the integration over the twistors associated to the loop) these terms give a contribution
\be
\begin{aligned}
	&\int\prod_{a=1}^4\frac{{\rm d}s_a}{s_a} \frac{{\rm d}u_1\,{\rm d}u_2}{(u_1-u_2)^2}\ 
	\Delta_\CSW(Z(s_1),Z(u_1))\ \Delta_\CSW(Z(s_2),Z(u_2))\ \Delta_\CSW(Z(s_3),Z(s_4))\\
	&=\ [\,*\,,i\!-\!1,i,A,\hat{B}]\,[\,*\,,j\!-\!1,j,A,\hat{\hat{B}}]\,[\,*\,,k\!-\!1,k,l\!-\!1,l]\,,
\label{NMHVloop2}
\end{aligned}
\ee
where 
\be
	\hat{B} = (AB)\cap(*,j\!-\!1,j)
	\qquad\hbox{and}\qquad 
	\hat{\hat{B}} = (AB)\cap(*,i\!-\!1,i)
\ee 
are the shifted loop variables.

If either $j=k$ or $l=i$ (or both) so that two of the fields are inserted on the same component of $C_n$, the corresponding  $\delbar^{-1}|_{C_n}$ propagators connect the fields directly rather than via a node, and consequently one must replace
\be
	\frac{{\rm d}s_3}{s_3}\to\frac{{\rm d}s_3}{(s_3-s_2)}
	\qquad\hbox{or}\qquad
	\frac{{\rm d}s_1}{s_1}\to\frac{{\rm d}s_1}{(s_4-s_1)}
\ee
in the measure of~\eqref{NMHVloop2}. These replacements lead to a corresponding shift in the external twistors  $Z_{j-1}=Z_{k-1}$ or $Z_{l-1}=Z_{i-1}$, given below. Once again adding up all the possible configurations of $\cA$ insertions one finds
\be
\begin{aligned}
	 &\left\la{\rm Tr}\left(\left.\delbar^{-1}\!\cA\,\delbar^{-1}\!\cA\,
	 \delbar^{-1}\!\cA\,\delbar^{-1}\!\cA\right|_{C_n}\right)\!\right\ra\\
	&\ =\ \int_\Gamma{\rm D}^{3|4}Z_A\wedge{\rm D}^{3|4}Z_B
	\ \sum\, [\,*\,,\widehat{i\!-\!1},i,A,\hat{B}]\, [\,*\,,j\!-\!1,j,A,\hat{\hat{B}}]\,[\,*\,,\widehat{k\!-\!1},k,l\!-\!1,l]\,,
\label{NMHVloopgeneric}
\end{aligned}
\ee
at order g$^2$, where the triple summation is over the range $1\leq i<j\leq k<l \leq i+n$, and where the hatted variables are defined as
\be
	\widehat{i-1} = 
		\begin{cases}
			(i\!-\!1,i)\cap(\,*\,,k\!-\!1,k) & \hbox{if } l=i+n\\
			\ i\!-\!1 & \hbox{otherwise,}
		\end{cases}
\ee
and
\be
	\widehat{k-1} =
		\begin{cases}
			(k\!-\!1,k)\cap(\,*\,,A,B) & \hbox{if } k=j\\
			\ k\!-\!1 & \hbox{otherwise,}
		\end{cases}
\ee
to account for the exceptional classes of diagram. Likewise, the shifted loop variables are here defined as
\be
	\hat{B} = (AB)\cap(\,*\,,j\!-\!1,j)\qquad\hbox{and}\qquad \hat{\hat{B}} = (AB)\cap(*,i\!-\!1,i)\, .
\ee
As promised, combining~\eqref{NMHVloopboundary} \&~\eqref{NMHVloopgeneric} gives the momentum twistor 1-loop NMHV amplitude, in the form presented in~\cite{CSWMat}.  There, it was proved analytically that~\eqref{NMHVloopboundary} +~\eqref{NMHVloopgeneric} agrees with the 1-loop NMHV integrand obtained from MHV rules in momentum space. That these expression in turn agree with conventional Feynman diagrams of the scattering amplitude was proved for any 1-loop amplitude in~\cite{Brandhuber:2005kd}. We  have thus proved
\be
	\big\la{\rm Tr}\left(\delbar^{-1}\!\cA\,\delbar^{-1}\!\cA\,\delbar^{-1}\!\cA\,\right)\!\big\ra_{\cO({\rm g}^2)}
	+\big\la{\rm Tr}\left(\delbar^{-1}\!\cA\,\delbar^{-1}\!\cA\,\delbar^{-1}\!\cA\,\delbar^{-1}\!\cA\,\right)
	\!\big\ra_{\cO({\rm g}^2)} = M_{\rm NMHV}^{(1)}\,.
\ee
Again, the space-time diagram for the correlation function is dual to the MHV diagram for the 1-loop NMHV scattering amplitude, as shown in figure~\ref{fig:1loopNMHVST}.

\FIGURE[t]{
	\includegraphics[width=100mm]{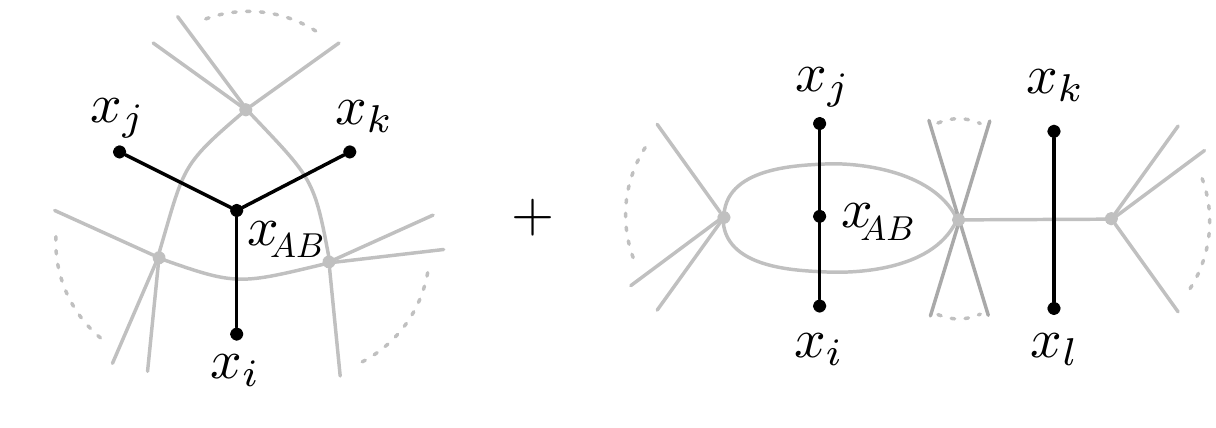}
	\caption{The two classes of MHV diagram contributing to the planar 1-loop NMHV amplitude are dual to the 
	diagrams of the two terms in~\eqref{1loopNMHVcorr} in the expansion of the Wilson loop.}
	\label{fig:1loopNMHVST}
}

\medskip

As at order g$^0$, notice that using the $\cN=0$ connection $\delbar+a$ to define the Wilson loop rather than the $\cN=4$ connection $\delbar+\cA$ would result in the vanishing of these contributions to the correlator. For the diagram shown in figure~\ref{fig:NMHVloopboundary}, this is because the $(a,g)$ propagator would have to link the three external $a$ fields to three $g$ fields on the line X, but being an MHV vertex, X only supports a most two $g$ fields. The contribution considered in~\eqref{NMHVloopgeneric} would also vanish for the same reason as at tree level. However, the diagram in figure~\ref{fig:1loopMHVtwistor} for the 1-loop MHV amplitude remains non-vanishing, because the two propagators perfectly match the two $g$ fields supported on X.


\subsection{The two loop MHV integrand}
\label{sec:2loopMHV}

The final example we shall consider in this paper is the integrand of the planar 2-loop, $n$ particle MHV amplitude. This arises from the same two terms
\be
	\left\la{\rm Tr}\left(\delbar^{-1}\!\cA\,\delbar^{-1}\!\cA\,\delbar^{-1}\!\cA\,\right)\right\ra
	+\left\la{\rm Tr}\left(\delbar^{-1}\!\cA\,\delbar^{-1}\!\cA\,\delbar^{-1}\!\cA\,\delbar^{-1}\!\cA\,\right)\right\ra
\ee	
as the 1-loop NMHV amplitude, but at order g$^4$. These diagrams thus involve two auxiliary twistor lines, $(A,B)$ and $(C,D)$.

Contributions from $\la{\rm Tr}\left(\delbar^{-1}\!\cA\,\delbar^{-1}\!\cA\,\delbar^{-1}\!\cA\,\right))\ra$ come from using twistor propagators to connect two of the field insertions to one of the auxiliary lines and the remaining field to the remaining auxiliary line, before finally connecting $(A,B)$ to $(C,D)$ with a further twistor propagator. The space-time diagram in shown in figure~\ref{fig:2loopMHV} and is again dual to one of the two classes of MHV diagram. Assuming the field insertions to be on the separate $C_n$ components $(i\!-\!1,i)$, $(j\!-\!1,j)$ and $(k\!-\!1,k)$ we find a contribution
\be
\begin{aligned}
	&\int_{(A,B)}\int_{(C,D)}
	\int\prod_{a=1}^3\frac{{\rm d}s_a}{s_a}\prod_{b=1}^3\frac{{\rm d}u_b}{(u_b-u_{b+1})}
	\prod_{c=1}^2\frac{{\rm d}v_c}{(v_c-v_{c+1})}\\
	&\hspace{1cm}\left\{\hspace{-0.4cm}\phantom{\int}\Delta_\CSW(Z(s_1),Z(u_1))\ \Delta_\CSW(Z(s_2),Z(u_2))\
	\Delta_\CSW(Z(u_3),Z(v_1))\ \Delta_\CSW(Z(v_2),Z(s_3))\right\}
\end{aligned}
\ee
where
\be
	Z(s_1) = s_1Z_{i-1}+Z_i\qquad Z(s_2)=s_2Z_{j-1}+Z_j\qquad Z(s_3)=s_3Z_{k-1}+Z_k
\ee
parametrize the locations of the external fields and for $b=1,2,3$ and $c=1,2$ we let
\be
	Z(u_b) = Z_A+u_bZ_B\qquad \hbox{and}\qquad Z(v_c)=Z_C+v_cZ_D
\ee
parametrize the locations that the propagators are attached to the auxiliary lines. We have used the shorthand $\int_{(A,B)}$ to indicate the integral $\int_\Gamma {\rm D}^{3|4}Z_A\wedge{\rm D}^{3|4}Z_B$. In the case that two external fields are inserted on the same component of $C_n$, say $(i\!-\!1,i) = (k\!-\!1,k)$, the different structure of the $\delbar^{-1}|_{C_n}$ operator with respect to the nodes leads to the replacement
\be
	\frac{{\rm d}s_1}{s_1} \ \to\ \frac{{\rm d}s_1}{(s_1-s_3)}
\ee
as in earlier examples. This again has the effect of shifting $Z_{i-1}$ in the arguments of the dual superconformal invariants. Summing over all possible insertions, we find that
\be
\begin{aligned}
	&\left\la{\rm Tr}\left(\delbar^{-1}\!\cA\,\delbar^{-1}\!\cA\,\delbar^{-1}\!\cA\,\right)\right\ra\ =\\
	&\quad\sum\,\int_{(A,B)}\int_{(C,D)}[*,\widehat{i-1},i,A,B']\,[*,j-1,j,A,B'']\,[*,A,B''',C,D'']\,[*,\widehat{k-1},k,C,D']\,,
\end{aligned}
\label{2loop1}
\ee
where the sum is over the range $1\leq i<j\leq k\leq i+n$ and the shifted loop and external twistors are
\be
\begin{aligned}
	&B' &= &\   (A,B)\cap(\,*\,,C,D) \qquad\qquad 	&D' &=\ (C,D)\cap(\,*\,,A,B)\\
	&B'' &= &\ (A,B)\cap(\,*\,,i\!-\!1,i) 				&D'' &= \ (C,D)\cap(\,*\,,k\!-\!1,k)\\
	&B''' &= &\  (A,B)\cap(\,*\,,j\!-\!1,j) & &
\end{aligned}
\ee
while the external shifts are
\be
\begin{aligned}
	\widehat{i\!-\!1} &= 
		\begin{cases}
			(i\!-\!1,i)\cap(\,*\,,C,D) 	& \hbox{if } k=i\\
			\ i\!-\!1 				& \hbox{otherwise}
		\end{cases}
	\\
	\widehat{k\!-\!1} &=
		\begin{cases}
			(k\!-\!1,k)\cap(\,*\,,A,B)	& \hbox{if } j=k\\
			\ k\!-\!1				& \hbox{otherwise.}
		\end{cases}
\end{aligned}
\ee

\FIGURE[t]{
	\includegraphics[height=40mm]{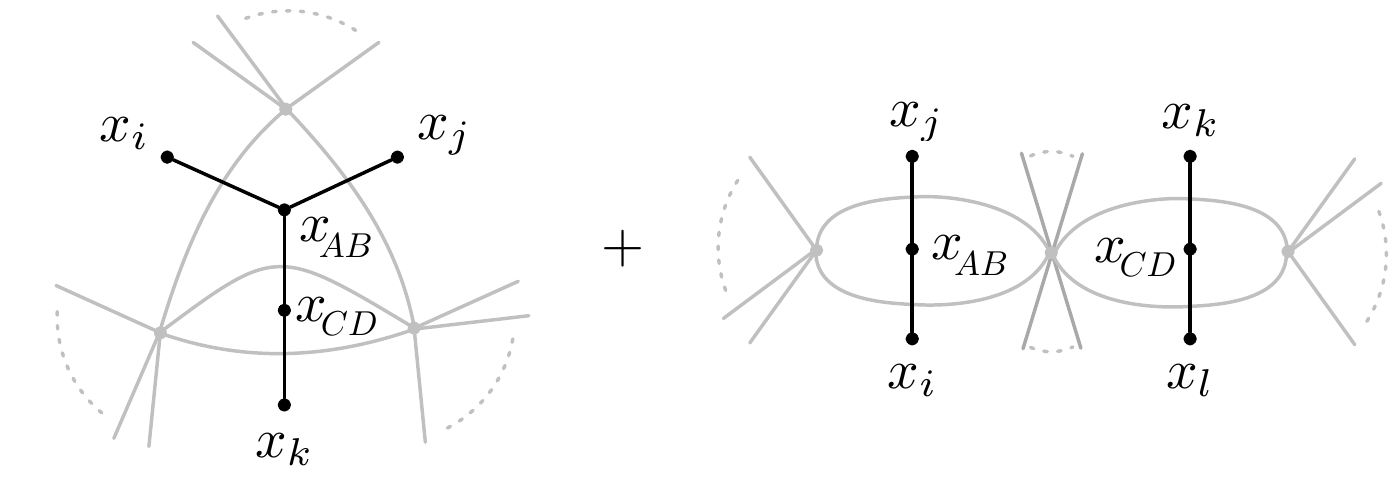}
	\caption{Diagrams contributing to the planar 2-loop MHV amplitude.}
	\label{fig:2loopMHV}
}

The remaining term $\la{\rm Tr}\left(\delbar^{-1}\!\cA\,\delbar^{-1}\!\cA\,\delbar^{-1}\!\cA\,\delbar^{-1}\!\cA\,\right)\ra$ is essentially just two copies of the 1-loop MHV expression, in that two of the fields are attached to the line $(A,B)$ while, quite independently, the other two are attached to $(C,D)$ (see figure~\ref{fig:2loopMHV} for the space-time MHV diagram). Therefore, in the case that all the external fields are inserted on different components of $C_n$, we obtain a contribution 
\be
	\int_{(A,B)}\hspace{-0.2cm}[\,\CSW\,,i\!-\!1,i,A,B']\, [\,\CSW\,j\!-\!1,j,A,B'']\,\times
	\int_{(C,D)}\hspace{-0.2cm}[\,\CSW\,,k\!-\!1,k,C,D']\,[\,\CSW\,,l\!-\!1,l,C,D'']
\ee
that is just the product of two independent copies of a 1-loop MHV contribution. The two terms mix when either $j=k$ or $l=i$, with the familiar effect that $Z_{k-1}$ and/or $Z_{i-1}$ become shifted. Summing over all possible insertion points leads to the order g$^4$ result
\be
\begin{aligned}
	&\left\la{\rm Tr}\left(\delbar^{-1}\!\cA\,\delbar^{-1}\!\cA\,\delbar^{-1}\!\cA\,\delbar^{-1}\!\cA\,\right)\right\ra\ =\\
	&\quad \sum\int_{(A,B)}\int_{(C,D)}
	[*,\widehat{i-1},i,A,B']\,[*,j-1,j,A,B'']\,[*,\widehat{k-1},k,C,D']\,[*,l-1,l,C,D'']\,,
\end{aligned}
\label{2loop2}
\ee
with the shifted loop variables
\be
\begin{aligned}
	&B' &= &\ (A,B)\cap(\,*\,,j\!-\!1,j) \qquad\qquad 	&D' &=\ (C,D)\cap(\,*\,,l\!-\!1,l)\\
	&B'' &= &\ (A,B)\cap(\,*\,,i\!-\!1,i) 				&D'' &=\ (C,D)\cap(\,*\,,k\!-\!1,k)
\end{aligned}
\ee
and external shifts
\be
\begin{aligned}
	\widehat{i\!-\!1} &=
		\begin{cases}
			(i\!-\!1,i)\cap(\,*\,,C,D) 	& \hbox{if } l=i\\
			\ i\!-\!1 				& \hbox{otherwise}\\
		\end{cases}
	\\
	\widehat{k\!-\!1} &=
		\begin{cases}
			(k\!-\!1,k)\cap(\,*\,,A,B) 	& \hbox{if } k=j\\
			\ k\!-\!1			 	& \hbox{otherwise.}
		\end{cases}
\end{aligned}
\ee
The summation range in this second term is $1\leq i<j\leq k< l\leq i$, again understood mod $n$.  Combining equations~\eqref{2loop1} \&~\eqref{2loop2} gives the complete integrand of the planar 2-loop $n$ particle MHV amplitude in the form found from the momentum twistor MHV rules~\cite{CSWMat}.


\section{The Correlation Function in Space-Time}
\label{sec:space-time}

Since our holomorphic Wilson-loop operator is essentially the twistor space transcription of the space-time Wilson loop, it is clear that if the space-time Wilson loop were defined fully supersymmetrically,  its correlation function would also give all N$^k$MHV amplitudes. The only question, therefore, is what the supersymmetric extension of the space-time Wilson loop should actually be.  Clearly, it must compute holonomy of some superconnection around a null polygon in superspace, with vertices at $(x_i,\theta_i)$.  In this section we see that the required superconnection is determined by a  supersymmetric version of the Penrose-Ward transform.  This can also be used to identify the space-time and twistor space definition of the Wilson loops. Note that the twistor action is most closely related to the space-time $\cN=4$ SYM action of Chalmers and Siegel~\cite{Chalmers:1996rq,Siegel:1992za}, so we expect that the resulting correlation function should be calculated using this action. We have yet to check whether the space-time correlator reproduces the scattering amplitudes.


\subsection{Construction of a space-time superconnection from the twistor data}

To construct a space-time superconnection from the twistor field $\cA$, we start from the $\cN=4$ generalisation of~\eqref{Hdef}, namely
\be
	\left.(\delbar+ \cA)\right|_{\rm X} {\rm H}(x,\theta,\lambda) = 0
\label{supersparl}
\ee
that follows because again the supertwistor bundle $E\to\CP^{3|4}$ is trivial when restricted to any $\CP^1$, so we can always find a holomorphic frame H on X and, for small enough $\cA$, we can find a family ${\rm H}(x,\theta,\lambda)$ of such frames simultaneously for every line X in some region of $\CP^{3|4}$.

In this equation, the restriction of $\Dbar$ to X is implemented by imposing the supersymmetric incidence relations
\be
\label{superincidence}
	\mu^{A'}=-\im x^{AA'}\lambda_A\,,\qquad \chi_a=\theta_a^A\lambda_A\,.
\ee
The remaining variables $(x,\theta,\lambda)$ are coordinates on the spin bundle $S$ and geometrically, the incidence relations define a projection $p:S\rightarrow \widehat U$ of the spin bundle to a region $\widehat U$ in twistor space (corresponding to a space-time region $U$). As the line ${\rm X}\subset\CP^{3|4}$ varies, restricting $\Dbar$ to X really means pulling back the twistor bundle $(E,\Dbar)$ to $S$ to obtain $p^*(E,\Dbar)$.  The solution $H$ to \eqref{supersparl} gives a frame of $p^*E$ that is holomorphic over the $\CP^1$ fibres.

The operators $\lambda^A\del/\del x^{AA'},\lambda^A\del/\del\theta^A_a , \del/\del\bar\lambda$ annihilate functions on $S$ that are the pullback of holomorphic functions on twistor space, since these depend on $(x,\theta)$ only via the incidence relations~\eqref{superincidence}. We can use the pullback of $\Dbar$ to extend these operators to the covariant operators
\be
	\cD := \left(\lambda^A\!\frac{\del}{\del x^{AA'}} +\cA_{A'}\,,\,
	\lambda^A\frac{\del}{\del\theta^{A}_{\, a}}\right)
\label{cDdef}
\ee
and
\be
	 \Dbar_{\bar\lambda }:=\frac \del{\del \bar \lambda }+ \cA_0
\ee
that act on sections and frames of $p^*E$.  In these operators, $\cA_{A'}$ and $\cA_0$ are the horizontal and vertical components of $p^*\cA$ with respect to the fibration $p:S\to \widehat{U}$. That is,
\be
	\cA_{A'}\equiv\lambda^A\frac{\del}{\del x^{AA'}}\lrcorner\, p^*\cA
	\qquad\hbox{and}\qquad
	\cA_0\equiv\frac \del{\del \bar \lambda }\, \lrcorner \, p^* \cA\,.
\ee
In terms of these operatosr,  \eqref{supersparl} is the equation $\Dbar_{\bar\lambda}{\rm H}=0$ on $S$.

Although $\{\lambda^A\del/\del x^{AA'},\lambda^A\del/\del\theta^A_a , \del/\del\bar\lambda\}$ commute and form an integrable distribution, the covariant operators do \emph{not} obey
\be
	[\cD_{A'},\cD_{B'}] =0, \qquad [\cD_{A'},\cD_{a}] = 0, \qquad [\cD_a,\cD_b]_+=0,\qquad
	[\cD,\Dbar_{\bar\lambda}]=0
\ee
unless the twistor $\Dbar$-operator itself obeys $\Dbar^2=0$. This condition would imply that $E$ was a holomorphic bundle of twistor space, corresponding to a self-dual connection on space-time as in the standard Penrose-Ward construction. Clearly, we do not wish to impose this in the off-shell context of a correlation function. However, without imposing any conditions,  $\Dbar$ does not give rise to a (super)connection on space-time at all, because it depends smoothly on six real variables rather than four.  The appropriate condition that gives an off-shell space-time connection is 
\be\label{curv-cond}
	[\cD, \Dbar_{\bar\lambda}]=0\, .
\ee
To see this, note that when~\eqref{curv-cond} holds we can act on~\eqref{supersparl} with $\cD$ to give
\be
	\Dbar_{\bar\lambda}(\cD {\rm H}) =0\, .
\ee
This in turn implies
\be
	 \bar\del ({\rm H}^{-1}\cD {\rm H})=0\, ,
\ee
where $\delbar$ is the d-bar operator acting on the $\lambda$s. Thus, although H$(x,\theta,\lambda)$ only depends smoothly on $\lambda$, the combination H$^{-1}\cD$H is in fact \emph{holomorphic}  globally on the $\lambda$-Riemann sphere.  Since it is clearly homogeneous of degree $+1$ (because $\cD$ is), an extension of Liouville's theorem says that ${\rm H}^{-1}\cD{\rm H}$ is in fact linear in $\lambda$, so that we have
\be
\label{construction-conn}
	{\rm H}^{-1}\cD {\rm H}=(\lambda^A A_{AA'}(x,\theta)\,,\, \lambda^A \Gamma_A^{\,a}(x,\theta))
\ee
where $ (A_{AA'}, \Gamma_A^{\,a})$ depend only on $(x,\theta)$. Thus we have constructed a superconnection
\be
	D_{AA'}=\frac\del{\del x^{AA'}} - A_{AA'}\, , \qquad D_A^a=\frac\del{\del \theta_a^A} -\Gamma^a_A
\ee 
on the bundle $E'$ over chiral super-space-time.

Rearranging~\eqref{construction-conn} and using the definition~\eqref{cDdef} we see that H$(x,\theta,\lambda)$ satisfies
\be\label{superconn}
	\lambda^A D_{AA'} {\rm H}=-\cA_{A'} {\rm H} 
	\qquad\hbox{and}\qquad
	\lambda^A D_A^a {\rm H} =0\, .
\ee
These equations imply that the space-time superconnection satisfies the integrability conditions
\be
	 \left[D_{(A}^{\ a},D_{B)B'}\right]=0\,, \qquad\qquad
	 \left[D_{(A}^{\ a},D_{B)}^{\ b}\right]_+=0\, .
\label{integrability}
\ee
While these conditions constrain the form of the component expansion of the superconnection, they do not imply the field equations, so the superconnection is off-shell.

Equation \eqref{superconn} also allows us to make contact with the parallel propagator along null geodesics and hence Wilson loops in Minkowski signature.  If $U$ is Minkowski space-time, then fibres of $S$ over twistor space are the null geodesics. Therefore any form on $S$ that is pulled back from a form on twistor space will vanish when contracted into  the vector field $\lambda^A\bar\lambda^{A'}\del_{AA'}$.  In particular $\bar\lambda^{A'}\cA_{A'}=0$, so that contracting the first equation in~\eqref{superconn} with $\bar\lambda^{A'}$ we find
\be 
	\lambda^A\bar\lambda^{A'}D_{AA'}{\rm H}=0\,, \qquad\qquad \lambda^A D_A^a {\rm H} =0\,.
\label{superparallel}
\ee
These equations uniquely fix H to be the parallel propagator along (super) null geodesics. Explicitly, if $\gamma:[0,1]\to U$ is a curve in chiral space-time with end points $\gamma(0) = (x_i,\theta_i)$ and $\gamma(1)=(x_{i+1},\theta_{i+1})$ and tangent vector
\be
	\gamma_*\left(\frac{\del}{\del t}\right) = \lambda_i^A\bar\lambda_i^{A'}\frac{\del}{\del x^{AA'}} 
	+ \lambda_i^A\eta_{ia}\frac{\del}{\del\theta^A_{\ a}}
\label{tangent}
\ee
($t\in[0,1]$ being a parameter along the interval), then integrating~\eqref{superparallel} gives
\be
	{\rm H}^{-1}(x_{i+1},\theta_{i+1};\lambda_i){\rm H}(x_i,\theta_i;\lambda_i) 
	= {\rm P}\exp\left(-\int_{[0,1]}\!\!\! \gamma^*\!\left(A_{AA'}{\rm d}x^{AA'}+\Gamma_{A}^{\ a}{\rm d}\theta^A_{\ a}\right)\right)
\label{superWLspacetime}
\ee
as a supersymmetrization of~\eqref{parallelpropZ}. Repeating for each leg of the Wilson loop will lead to the space-time formulation of the supersymmetric Wilson loop defined in~\eqref{WLtws} for twistor space\footnote{When $U$ is a real 4-dimensional slice of complex conformal Minkowski space that admits no null tangent vectors, $\widehat U$ has 6 real  bosonic dimensions in twistor space and the condition~\eqref{curv-cond} leads to the vanishing of two of the three components of $\Dbar^2$. The reconstruction of an off-shell connection is then straightforward.  However, when $U$ is a real slice of Lorentz signature, $\widehat U$ degenerates to become a 5-dimensional CR manifold, $PN$, defined by $Z\cdot \bar Z=0$.  On such a CR manifold, $\Dbar^2$ has only one component and so~\eqref{curv-cond} implies $\Dbar^2=0$.  Thus, if $\Dbar$ defines a connection on space-time, it is necessarily self-dual.  In order to obtain the above identification of twistor space with space-time Wilson loops for a completely off-shell connection, in this situation we must use a limiting argument from slices with no null tangent vectors analogous to standard arguments for obtaining correlation functions for operators inserted at light-like separated points obtained as limits of those inserted at space-like separated points.}. We  expect that the correlation function of the space-time supersymmetric Wilson loop arising from~\eqref{superWLspacetime} also computes the complete planar S-matrix of $\cN=4$ SYM.

\FIGURE[t]{
	\includegraphics[height=50mm]{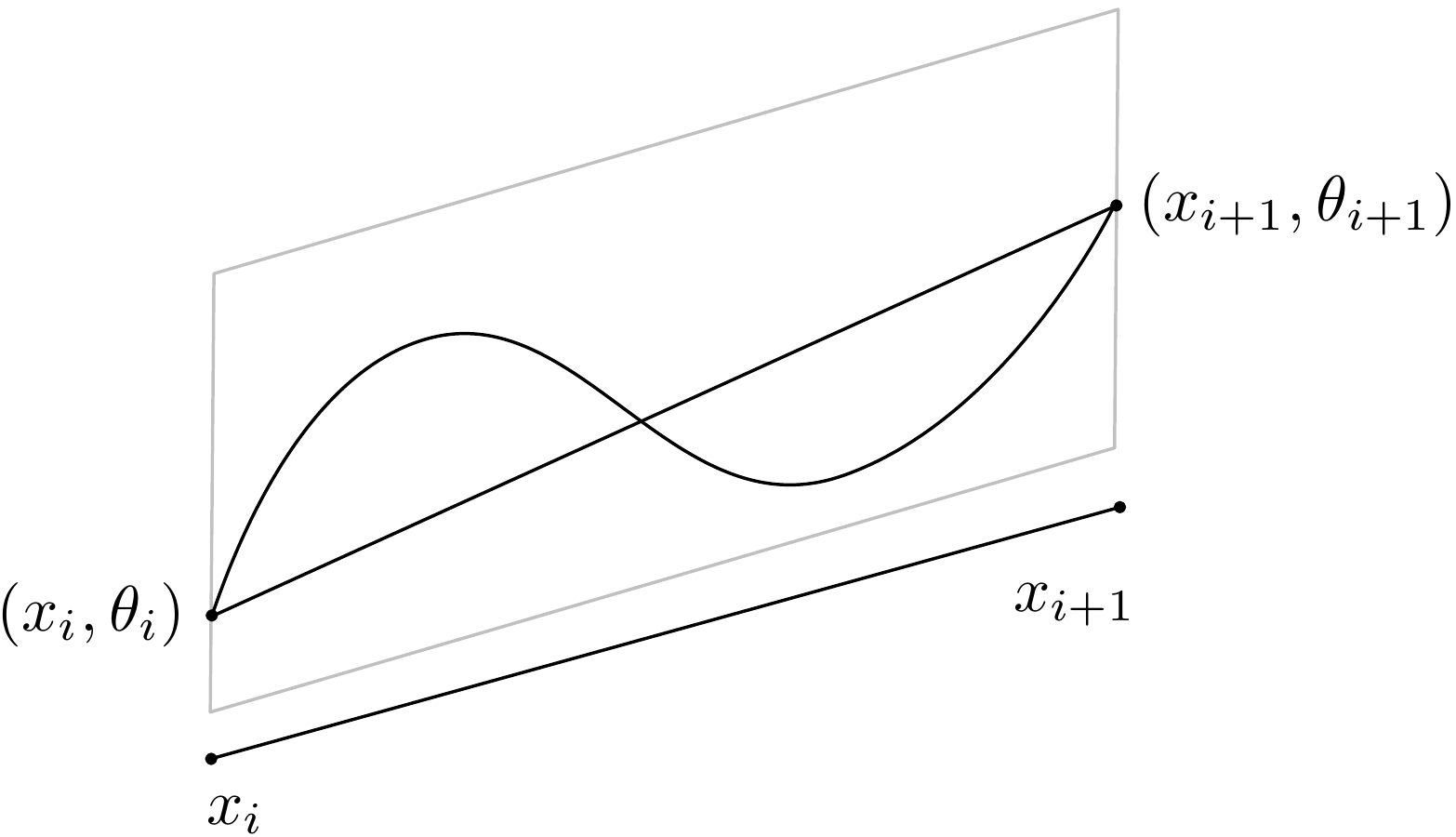}
	\caption{The integrablity conditions~\eqref{integrability} imply that the curvature of the space-time superconnection 		vanishes above any null geodesic in the non-supersymmetric space-time. The Wilson loop is thus independent
		of the choice of lift of this null geodesic, provided its has the same endpoints.}
	\label{fig:superlift}
}

Note that although the relation to scattering amplitudes provides us with a preferred tangent vector~\eqref{tangent}, the integrability conditions
\be
	 \lambda^A\lambda^B \left[D_A^{\ a},D_{BB'}\right]=0\,, \qquad\qquad
	 \lambda^A\lambda^B \left[D_A^{\ a},D_B^{\ b}\right]_+=0
\ee
of equation~\eqref{integrability} imply that the curvature of the superconnection vanishes provided we remain above any null ray in the bosonic space-time. Therefore, at least at the classical level, the parallel propagator~\eqref{superWLspacetime} is independent of the choice of lift of the null geodesic $\lambda_i\bar\lambda_i\del /\del x$ to superspace (see figure~\ref{fig:superlift}). Said differently, there exists a gauge for $(A,\Gamma)$ in which the dependence of~\eqref{superWLspacetime} on the Grassmann coordinates $\theta$ comes only from the endpoints.


\subsection{Component form of the space-time superconnection}

Finally, as an exercise we briefly indicate how the component expansion of this space-time superconnection is related to the component expansion of the twistor $\cN=4$ superfield. This is again determined by the integrability conditions~\eqref{integrability} and may be obtained explicitly be expanding out all the fermionic coordinates at each stage in the above procedure.  We calculate the superconnection to order $\theta^2$ in the general case, and give the full answer at the end for the abelian case.

The twistor superfield is
\be
\label{a-expand}
	\cA(Z,\chi) = a(Z) + \chi_a\,\gamma^a(Z)+\frac{1}{2}\chi_a\chi_b\,\phi^{ab}(Z) 
	+ \frac{\epsilon^{abcd}}{3!}\chi_a\chi_b\chi_c\,\tilde\gamma_d 
	+ \frac{\epsilon^{abcd}}{4!}\chi_a\chi_b\chi_c\chi_d\, g(Z)\,,
\ee
where the component of $(\chi)^r $ is a (0,1)-form on the appropriate region $U\subset\CP^3$, homogeneous of degree $-r$. Under the Penrose-Ward transform, $\gamma^a(Z)$ and $\tilde\gamma_a(Z)$ correspond respectively to positive and negative helicity gluinos, $\phi^{ab}(Z)$ to  scalars, and $g(Z)$ to the anti-self-dual part of the Yang-Mills field. We can similarly expand equation~\eqref{supersparl} in the fermionic coordinates $\theta$ and solve term by term.  Set
\be 
	{\rm H}(x,\theta,\lambda)=H(x,\lambda)+\theta_a^Ah^a_A(x,\lambda)+\theta_a^A\theta_b^Bh^{ab}_{AB}(x,\lambda)
	+\ldots +\theta^8h_8(x,\lambda)
\ee
and expand out equation~\eqref{supersparl} using this, \eqref{a-expand} and \eqref{superincidence} for $\chi^a$ to get the component equations
\be
\begin{aligned}
\label{component-sparl} 
	0&= (\delbar +a)|_{\rm X} H(x,\lambda)\\ 
	0&= (\delbar+a)|_{\rm  X} h_A^a(x,\lambda)+\lambda_A\,\gamma^a(x,\lambda) H(x,\lambda)\\ 
	0&=(\delbar+a)|_{\rm X} h_{AB}^{ab}(x,\lambda)+\lambda_A\,\gamma^a(x,\lambda) h_B^b(x,\lambda) 
	+\lambda_A\lambda_B\, \phi^{ab}(x,\lambda)H(x,\lambda)\\
	&\qquad\qquad\vdots
\end{aligned}
\ee
where the last equation is understood to have the symmetries of $\theta^A_a\theta^B_b$ imposed on it.

The leading order part is the non-supersymmetric form~\eqref{Hdef} that fixes the holomorphic frame $H(x,\lambda)$, while  the subsequent terms are what one expects for the Penrose transform for fields that are background coupled to the
connection arising from the first term. We can solve these component equations hierarchically. The first leads to the ordinary space-time connection $A(x)=A(x,\theta)|_{\theta=0}$, because we use the bosonic part of~\eqref{construction-conn} at zeroth order in $\theta$ to fix $A(x)$ in  terms of $H$ as
\be
	\lambda^AA_{AA'}(x)= \lambda^AH^{-1}\del_{AA'} H\, .
\ee

The subleading parts can be simplified by using $H$ as a gauge transformation to replace the $\delbar+a|{\rm X}$ operator simply by $\delbar$.   The second equation of~\eqref{component-sparl} becomes
\be
\label{sublead-sparl}
	\delbar ( H^{-1} h_A^a(x,\lambda))+\lambda_AH^{-1}\gamma^a H=0\, ,
\ee
where here and in what follows, $\delbar$ always refers to the d-bar operator on X. The related equation
\be
	\delbar  \psi^a +H^{-1}\gamma^a(x,\lambda)H=0
\ee
has a unique solution $\psi^a$ that has weight $-1$ in $\lambda$. The quantity 
\be 
\Psi^a_{A'} := \lambda^A D_{AA'}\psi^a
\ee
is holomorphic in $\lambda$ because $\lambda^AD_{AA'}$ annihilates $H^{-1}\gamma^aH$,  and since $\Psi$ has weight zero in $\lambda$, Liouville's theorem states that it is in fact independent of $\lambda$.
Therefore, setting
\be
	H^{-1} \, h_A^a= \psi^a \lambda_A
\ee
and using~\eqref{construction-conn}  we determine the coefficient of $\theta$ in the expansion of the connection to be:
\be
	A_{AA'}(x,\theta)=A_{AA'}(x) + \theta_{Aa}\Psi^a_{A'}(x) + O(\theta^2) \, , \qquad   \Gamma^a_A= 0+ O(\theta)
\ee
Note that we are free to add a function $\rho_A^a(x)$ that is independent of $\lambda$  to the solution of~\eqref{sublead-sparl}. Doing so leads to the addition of  a term $D_{AA'}(\rho^b_B \theta_b^B)$ to $A_{AA'}$ and $\rho^a_A$ to $\Gamma^a_A$, representing a supersymmetric gauge transformation.

We can proceed similarly with the third equation in~\eqref{component-sparl}. Using the solutions of the first two equations in~\eqref{component-sparl}, we may write the third as
\be
\begin{aligned}
	0&=\delbar (H^{-1}h^{ab}_{AB}(x,\lambda)) + \lambda_A(H^{-1}\gamma^a H)(H^{-1} h_B^b) +
	\lambda_A\lambda_BH^{-1}\phi^{ab}H \\
	&=\delbar (H^{-1}h^{ab}_{AB}(x,\lambda)) - \lambda_A\lambda_B(\delbar\psi^a)\psi^b + 
	\lambda_A\lambda_BH^{-1}\phi^{ab}H \,.
\end{aligned}
\ee
As with the gluino, we first consider the related equation
\be
	\delbar (H^{-1}h^{ab}_{A}(x,\lambda)) + \lambda_A(\delbar \psi^a)\psi^b +	\lambda_AH^{-1}\phi^{ab}H =0
\ee
which has a unique soluton $H^{-1}h^{ab}_{A}=m^{ab}_A(x,\lambda)$ of homogeneity $-1$ in $\lambda$. We can now observe that 
\be
\Phi^{ab}:=\lambda^Am^{ab}_A
\ee
has homogeneity degree zero and is holomorphic on the sphere and so depends only on $x$. We identify it with the scalar field in the multiplet.   Setting $H^{-1}h^{ab}_{AB}=m^{ab}_{(A}\lambda_{B)}$  we now observe that 
\be
\bar\del (\lambda^BD_{BA'} m_A^{ab}+\lambda_A\psi^{[a}\Psi^{b]}_{A'})=0
\ee
so by a now familiar argument 
\be
\Phi^{ab}_{AA'}:=\lambda^BD_{BA'} m_A^{ab}+\lambda_A\psi^{[a}\Psi^{b]}_{A'}
\ee
depends only on $x$ and by contraction with $\lambda^A$ we discover that
\be
\Phi_{AA'}^{ab}=D_{AA'}\Phi^{ab}
\ee
and we find from equation~\eqref{construction-conn} that
\be
A_{AA'BC}^{ab}=\varepsilon_{A(B}D_{C)A'}\Phi^{ab}
\ee  
is the coefficient of $\theta^2$ in the expansion of $A(x,\theta)$. Therefore, to second order in $\theta$, we have the superconnection
\be
\begin{aligned}
	A_{AA'}(x,\theta)&=A_{AA'}(x) + \theta_{Aa}\Psi^a_{A'}(x) + \theta_{Aa}\theta^B_{\ b}D_{BA'}\Phi^{ab} + O(\theta^3)\\
	\Gamma^a_{\,A}(x,\theta) &=\theta_{Ab}\Phi^{ab} + O(\theta^2)
\end{aligned}	
\ee	
Proceeding in a similar way (see also~\cite{Sparling}) will lead to a complete component expansion of the space-time superconnection in terms of (derivatives of) the component fields.   This is easily performed in the abelian case where we obtain
\be
\begin{aligned}
	A_{AA'}(x,\theta)&=A_{AA'}(x) + \Psi^a_{A'}(x)\theta_{Aa} + D_{BA'}\Phi^{ab} \theta_{Aab}^{2B} +  D_{BA'}\widetilde\Psi_{Cc}\theta^{3BCc}_A + D_{BA'}G_{CD}\theta^{4BCD}_A\\
	\Gamma^a_{\,A}(x,\theta) &=\Phi^{ab}\theta_{Ab} + \Psi_{B}^{abc}\theta^{2B}_{Abc} + G_{BC}\theta^{3BCa}_A \, 
\end{aligned}	
\ee	
where we have introduced the notation
\be
\theta^{2AB}_{ab}=\theta^{(A}_a\theta^{B)}_b\, , \qquad \theta^{3ABCa}=\varepsilon ^{abcd}\theta_b^A\theta_c^B\theta_d^C\, ,\qquad \theta^{4ABCD}=\varepsilon ^{abcd}\theta_a^A\theta_b^B\theta_c^C\theta_d^D\, 
\ee 
(all symmetric on their spinor indices).

\acknowledgments

It is a pleasure to thank Tim Adamo, Nima Arkani-Hamed, Rutger Boels, Jacob Bourjaily, Mathew Bullimore, Freddy Cachazo, Simon Caron-Huot, James Drummond and Johannes Henn for many helpful discussions and much inspiration. While in the later stages of writing this paper, we learnt of related independent work by Simon Caron-Huot based on a space-time approach. The work of DS is supported by the Perimeter Institute for Theoretical Physics. Research at the Perimeter Institute is supported by the Government of Canada through Industry Canada and by the Province of Ontario through the Ministry of Research $\&$ Innovation. The work of LM  was financed in part by EPSRC grant number EP/F016654.


\bibliographystyle{JHEP}
\bibliography{WilsonLoop}

\providecommand{\href}[2]{#2}\begingroup\raggedright\begin{thebibliography}{10}

\bibitem{CSWMat}
M.~Bullimore, L.~Mason, and D.~Skinner, {\it {MHV Diagrams and BCFW Diagrams in
  Momentum Twistor Space}},  \href{http://arxiv.org/abs/1009.1854}{{\tt
  arXiv:1009.1854}}.

\bibitem{HuggettTod}
S.~Huggett and P.~Tod, {\em {An Introduction to Twistor Theory}}.
\newblock Student Texts 4. London Mathematical Society, 1985.

\bibitem{Penrose:1986ca}
R.~Penrose and W.~Rindler, {\em Spinors and Space-Time}, vol.~2.
\newblock Cambridge University Press, 1986.

\bibitem{WardWells}
R.~Ward and R.~Wells, {\em {Twistor Geometry and Field Theory}}.
\newblock CUP, 1990.

\bibitem{Drummond:2008bq}
J.~M. Drummond, J.~Henn, G.~P. Korchemsky, and E.~Sokatchev, {\it {Generalized
  Unitarity for $\cN=4$ Super-Amplitudes}},
  \href{http://arxiv.org/abs/0808.0491}{{\tt arXiv:0808.0491}}.

\bibitem{Hodges:2009hk}
A.~Hodges, {\it {Eliminating Spurious Poles from Gauge-Theoretic Amplitudes}},
  \href{http://arxiv.org/abs/0905.1473 [hep-th]}{{\tt arXiv:0905.1473
  [hep-th]}}.

\bibitem{Mason:2009qx}
L.~Mason and D.~Skinner, {\it {Dual Superconformal Invariance, Momentum
  Twistors and Grassmannians}},  {\em JHEP} {\bf 11} (2009) 045,
  [\href{http://arxiv.org/abs/0909.0250}{{\tt arXiv:0909.0250}}].

\bibitem{Drummond:2007aua}
J.~Drummond, G.~Korchemsky, and E.~Sokatchev, {\it {Conformal Properties of
  Four-Gluon Planar Amplitudes and Wilson Loops}},  {\em Nucl. Phys.} {\bf
  B795} (2007) 385--408, [\href{http://arxiv.org/abs/0707.0243 [hep-th]}{{\tt
  arXiv:0707.0243 [hep-th]}}].

\bibitem{Drummond:2007au}
J.~Drummond, J.~Henn, G.~Korchemsky, and E.~Sokatchev, {\it {Conformal Ward
  identities for Wilson loops and a test of the duality with gluon
  amplitudes}},  {\em Nucl. Phys.} {\bf B826} (2010) 337--364,
  [\href{http://arxiv.org/abs/0812.1223 [hep-th]}{{\tt arXiv:0812.1223
  [hep-th]}}].

\bibitem{Drummond:2008aq}
J.~Drummond, J.~Henn, G.~Korchemsky, and E.~Sokatchev, {\it {Hexagon Wilson
  Loop = Six-Gluon MHV Amplitude}},  {\em Nucl. Phys.} {\bf B815} (2009)
  142--173, [\href{http://arxiv.org/abs/0803.1466 [hep-th]}{{\tt
  arXiv:0803.1466 [hep-th]}}].

\bibitem{Brandhuber:2007yx}
A.~Brandhuber, P.~Heslop, and G.~Travaglini, {\it {MHV Amplitudes in $\cN=4$
  Super Yang-Mills and Wilson Loops}},  {\em Nucl. Phys.} {\bf B794} (2008)
  231--243, [\href{http://arxiv.org/abs/0707.1153}{{\tt arXiv:0707.1153}}].

\bibitem{Anastasiou:2009kna}
C.~Anastasiou, A.~Brandhuber, P.~Heslop, V.~Khoze, W.~Spence, and
  G.~Travaglini, {\it {Two-Loop Polygon Wilson Loops in $\cN=4$ SYM}},  {\em
  JHEP} {\bf 05} (2009) 115, [\href{http://arxiv.org/abs/0902.2245
  [hep-th]}{{\tt arXiv:0902.2245 [hep-th]}}].

\bibitem{DelDuca:2009au}
V.~Del~Duca, C.~Duhr, and V.~Smirnov, {\it {An Analytic Result for the Two-Loop
  Hexagon Wilson Loop in $\cN=4$ SYM}},  {\em JHEP} {\bf 03} (2010) 099,
  [\href{http://arxiv.org/abs/0911.5332}{{\tt arXiv:0911.5332}}].

\bibitem{DelDuca:2010zg}
V.~Del~Duca, C.~Duhr, and V.~Smirnov, {\it {The Two-Loop Hexagon Wilson Loop in
  $\cN=4$ SYM}},  {\em JHEP} {\bf 05} (2010) 084,
  [\href{http://arxiv.org/abs/1003.1702}{{\tt arXiv:1003.1702}}].

\bibitem{DelDuca:2010zp}
V.~Del~Duca, C.~Duhr, and V.~Smirnov, {\it {A Two-Loop Octagon Wilson Loop in
  $\cN=4$ SYM}},  \href{http://arxiv.org/abs/1006.4127 [hep-th]}{{\tt
  arXiv:1006.4127 [hep-th]}}.

\bibitem{Goncharov:2010jf}
A.~Goncharov, M.~Spradlin, C.~Vergu, and A.~Volovich, {\it {Classical
  Polylogarithms for Amplitudes and Wilson Loops}},
  \href{http://arxiv.org/abs/1006.5703 [hep-th]}{{\tt arXiv:1006.5703
  [hep-th]}}.

\bibitem{Alday:2007hr}
L.~F. Alday and J.~Maldacena, {\it {Gluon Scattering Amplitudes at Strong
  Coupling}},  {\em JHEP} {\bf 06} (2007) 064,
  [\href{http://arxiv.org/abs/arxiv 0705.0303 [hep-th]}{{\tt arxiv 0705.0303
  [hep-th]}}].

\bibitem{Alday:2009ga}
L.~F. Alday and J.~Maldacena, {\it {Minimal Surfaces in AdS and the Eight Gluon
  Scattering Amplitude at Strong Coupling}},
  \href{http://arxiv.org/abs/0903.4707 [hep-th]}{{\tt arXiv:0903.4707
  [hep-th]}}.

\bibitem{Alday:2009dv}
L.~F. Alday, D.~Gaiotto, and J.~Maldacena, {\it {Thermodynamic Bubble Ansatz}},
   \href{http://arxiv.org/abs/0911.4709 [hep-th]}{{\tt arXiv:0911.4709
  [hep-th]}}.

\bibitem{Alday:2010vh}
L.~F. Alday, J.~Maldacena, A.~Sever, and P.~Vieira, {\it {Y-system for
  Scattering Amplitudes}},  \href{http://arxiv.org/abs/1002.2459 [hep-th]}{{\tt
  arXiv:1002.2459 [hep-th]}}.

\bibitem{Mason:2005zm}
L.~Mason, {\it {Twistor Actions for Non Self-Dual Fields}},  {\em JHEP} {\bf
  10} (2005) 009, [\href{http://arxiv.org/abs/hep-th/0507269}{{\tt
  hep-th/0507269}}].

\bibitem{Boels:2006ir}
R.~Boels, L.~Mason, and D.~Skinner, {\it {Supersymmetric Gauge Theories in
  Twistor Space}},  {\em JHEP} {\bf 02} (2007) 014,
  [\href{http://arxiv.org/abs/hep-th/0604040}{{\tt hep-th/0604040}}].

\bibitem{Boels:2007qn}
R.~Boels, L.~Mason, and D.~Skinner, {\it {From Twistor Actions to MHV
  Diagrams}},  {\em Phys. Lett.} {\bf B648} (2007) 90--96,
  [\href{http://arxiv.org/abs/hep-th/0702035}{{\tt hep-th/0702035}}].

\bibitem{Cachazo:2004kj}
F.~Cachazo, P.~Svrcek, and E.~Witten, {\it {MHV Vertices and Tree Amplitudes in
  Gauge Theory}},  {\em JHEP} {\bf 09} (2004) 006,
  [\href{http://arxiv.org/abs/hep-th/0403047}{{\tt hep-th/0403047}}].

\bibitem{Bern:2008ap}
Z.~Bern {\em et~al.}, {\it {The Two-Loop Six-Gluon MHV Amplitude in Maximally
  Supersymmetric Yang-Mills Theory}},  {\em Phys. Rev.} {\bf D78} (2008)
  045007, [\href{http://arxiv.org/abs/0803.1465}{{\tt arXiv:0803.1465}}].

\bibitem{Hodges:2010kq}
A.~Hodges, {\it {The Box Integrals in Momentum Twistor Geometry}},
  \href{http://arxiv.org/abs/1004.3323}{{\tt arXiv:1004.3323}}.

\bibitem{Mason:2010pg}
L.~Mason and D.~Skinner, {\it {Amplitudes at Weak Coupling as Polytopes in
  AdS$_5$}},  \href{http://arxiv.org/abs/1004.3498}{{\tt arXiv:1004.3498}}.

\bibitem{Drummond:2010mb}
J.~Drummond and J.~Henn, {\it {Simple Loop Integrals and Amplitudes in $\cN=4$
  SYM}},  \href{http://arxiv.org/abs/1008.2965}{{\tt arXiv:1008.2965}}.

\bibitem{Alday:2010jz}
L.~F. Alday, {\it {Some Analytic Results for Two-Loop Scattering Amplitudes}},
  \href{http://arxiv.org/abs/1009.1110}{{\tt arXiv:1009.1110}}.

\bibitem{Alday:2009zm}
L.~F. Alday, J.~Henn, J.~Plefka, and T.~Schuster, {\it {Scattering Into the
  Fifth Dimension of $\mathcal{N}=4$ Super Yang-Mills}},  {\em JHEP} {\bf 01}
  (2010) 077, [\href{http://arxiv.org/abs/0908.0684 hep-th}{{\tt
  arXiv:0908.0684 hep-th}}].

\bibitem{Arkani-Hamed:2010kv}
N.~Arkani-Hamed, J.~Bourjaily, F.~Cachazo, S.~Caron-Huot, and J.~Trnka, {\it
  {The All-Loop Integrand for Scattering Amplitudes in Planar $\cN=4$ SYM}},
  \href{http://arxiv.org/abs/1008.2958 [hep-th]}{{\tt arXiv:1008.2958
  [hep-th]}}.

\bibitem{Drummond:2009fd}
J.~Drummond, J.~Henn, and J.~Plefka, {\it {Yangian Symmetry of Scattering
  Amplitudes in $\mathcal{N}=4$ Super Yang-Mills}},  {\em JHEP} {\bf 05} (2009)
  046, [\href{http://arxiv.org/abs/0902.2987 [hep-th]}{{\tt arXiv:0902.2987
  [hep-th]}}].

\bibitem{Chalmers:1996rq}
G.~Chalmers and W.~Siegel, {\it {The Self-Dual Sector of {QCD} Amplitudes}},
  {\em Phys. Rev.} {\bf D54} (1996) 7628--7633,
  [\href{http://arxiv.org/abs/hep-th/9606061}{{\tt hep-th/9606061}}].

\bibitem{Siegel:1992za}
W.~Siegel, {\it {The $\cN=2(4)$ String is Self-dual $\cN=4$ Yang-Mills}},
  \href{http://arxiv.org/abs/hep-th/9205075}{{\tt hep-th/9205075}}.

\bibitem{Alday:2010zy}
L.~F. Alday, B.~Eden, G.~Korchemsky, J.~Maldacena, and E.~Sokatchev, {\it {From
  Correlation Functions to Wilson Loops}},
  \href{http://arxiv.org/abs/1007.3243 [hep-th]}{{\tt arXiv:1007.3243
  [hep-th]}}.

\bibitem{Eden:2010zz}
B.~Eden, G.~Korchemsky, and E.~Sokatchev, {\it {From Correlation Functions to
  Scattering Amplitudes}},  \href{http://arxiv.org/abs/1007.3246 [hep-th]}{{\tt
  arXiv:1007.3246 [hep-th]}}.

\bibitem{Eden:2010ce}
B.~Eden, G.~Korchemsky, and E.~Sokatchev, {\it {More on the Duality Correlators
  / Amplitudes}},  \href{http://arxiv.org/abs/1009.2488}{{\tt
  arXiv:1009.2488}}.

\bibitem{withMat}
T.~Adamo, M.~Bullimore, L.~Mason, and D.~Skinner, {\it {To appear}},  2010.

\bibitem{Witten:1992fb}
E.~Witten, {\it {Chern-Simons Gauge Theory as a String Theory}},  {\em
  Prog.~Math.} {\bf 133} (1995) 637--678,
  [\href{http://arxiv.org/abs/hep-th/9207094}{{\tt hep-th/9207094}}].

\bibitem{Witten:2003nn}
E.~Witten, {\it {Perturbative Gauge Theory as a String Theory in Twistor
  Space}},  {\em Commun. Math. Phys.} {\bf 252} (2004) 189--258,
  [\href{http://arxiv.org/abs/hep-th/0312171}{{\tt hep-th/0312171}}].

\bibitem{QuillenLine}
D.~Quillen, {\it {Determinants of Cauchy-Riemann Operators over Riemann
  Surfaces}},  {\em Func.~Anal.~Appl.} {\bf 19} (1985) 37--41.

\bibitem{Ward:1977ta}
R.~Ward, {\it {On Self-Dual Gauge Fields}},  {\em Phys. Lett.} {\bf A61} (1977)
  81--82.

\bibitem{Lovelace:2010ev}
C.~Lovelace, {\it {Twistors Versus Harmonics}},
  \href{http://arxiv.org/abs/1006.4289}{{\tt arXiv:1006.4289}}.

\bibitem{Henn:2010bk}
J.~Henn, S.~Naculich, H.~Schnitzer, and M.~Spradlin, {\it {Higgs-regularized
  three-loop four-gluon amplitude in $\cN=4$ SYM: exponentiation and Regge
  limits}},  \href{http://arxiv.org/abs/1001.1358}{{\tt arXiv:1001.1358}}.

\bibitem{Bern:2009xq}
Z.~Bern, J.~J. Carrasco, H.~Ita, H.~Johansson, and R.~Roiban, {\it {On the
  Structure of Supersymmetric Sums in Multi-Loop Unitarity Cuts}},  {\em Phys.
  Rev.} {\bf D80} (2009), no.~065029
  [\href{http://arxiv.org/abs/0903.5348}{{\tt arXiv:0903.5348}}].

\bibitem{Sever:2009aa}
A.~Sever and P.~Vieira, {\it {Symmetries of the $\cN=4$ SYM S-matrix}},
  \href{http://arxiv.org/abs/0908.2437}{{\tt arXiv:0908.2437}}.

\bibitem{Mason:2009sa}
L.~Mason and D.~Skinner, {\it {Scattering Amplitudes and BCFW Recursion in
  Twistor Space}},  \href{http://arxiv.org/abs/0903.2083}{{\tt
  arXiv:0903.2083}}.

\bibitem{Atiyah:1981ey}
M.~F. Atiyah, {\it {Green's Functions for Self-Dual Four Manifolds}},  {\em
  Adv. Math. Supp.} {\bf 7A} (1981) 129--158.

\bibitem{Adamo:2010}
T.~Adamo, L.~Mason, and D.~Skinner, {\it {To appear}},  2010.

\bibitem{Drummond:2008vq}
J.~M. Drummond, J.~Henn, G.~P. Korchemsky, and E.~Sokatchev, {\it {Dual
  Superconformal Symmetry of Scattering Amplitudes in $\cN=4$ Super-Yang-Mills
  Theory}},  \href{http://arxiv.org/abs/0807.1095}{{\tt arXiv:0807.1095}}.

\bibitem{Brandhuber:2004yw}
A.~Brandhuber, W.~Spence, and G.~Travaglini, {\it {One-loop Gauge Theory
  Amplitudes in $\cN=4$ Super Yang-Mills from MHV Vertices}},  {\em Nucl.
  Phys.} {\bf B706} (2005) 150--180,
  [\href{http://arxiv.org/abs/hep-th/0407214}{{\tt hep-th/0407214}}].

\bibitem{Bena:2004xu}
I.~Bena, Z.~Bern, and D.~Kosower, {\it {Loops in Twistor Space}},  {\em Phys.
  Rev.} {\bf D71} (2005) 106010,
  [\href{http://arxiv.org/abs/hep-th/0410054}{{\tt hep-th/0410054}}].

\bibitem{Brandhuber:2005kd}
A.~Brandhuber, W.~Spence, and G.~Travaglini, {\it {From Trees to Loops and
  Back}},  {\em JHEP} {\bf 01} (2006) 142,
  [\href{http://arxiv.org/abs/hep-th/0510253}{{\tt hep-th/0510253}}].

\bibitem{Sparling}
G.~Sparling, {\it {Dynamically Broken Symmetry and Global Yang-Mills in
  Minkowski Space}},  in {\em Further Advances in Twistor Theory}, vol.~1.
\newblock Longman Scientific and Technical, 1977.

\end{thebibliography}\endgroup

\end{document}